%% file: limited-look-ahead.tex
\documentclass[]{article}

\usepackage{lineno,hyperref}
\usepackage[margin=1in]{geometry}

\usepackage{chrkroer}

\usepackage[ruled,noend]{algorithm2e}

\usepackage{paralist}
\usepackage{multicol}
\usepackage{amsfonts}

\usepackage{subfig}
\usepackage{tikz}
\usetikzlibrary{fit,arrows,calc,positioning,shapes.geometric}
\usepackage{tikz-qtree}
\usepackage{standalone}
\usepackage{natbib}

\newcommand{\tuple}[1]{\ensuremath{\left \langle #1 \right \rangle }}
\newcommand{\child}[2]{\ensuremath{t^{#1}_{#2}}}

\newcommand{\KJ}{KJ}
\renewcommand{\la}{\ensuremath{l}}
\newcommand{\rational}{\ensuremath{r}}

 \title{Limited Lookahead in Imperfect-Information Games}
\author{Christian Kroer\\
  IEOR Department\\
  Columbia University\\
  New York City, NY, USA\\
  \texttt{christian.kroer@columbia.edu}
  \and
Tuomas Sandholm \\
Computer Science Department\\ Carnegie Mellon University\\
Strategy Robot, Inc.\\
Strategic Machine, Inc.\\
Optimized Markets, Inc.\\
Pittsburgh, PA, USA\\
\texttt{sandholm@cs.cmu.edu}
}

\begin{document}
\maketitle

% \tnotetext[mytitlenote]{A short preliminary version of this paper appeared in the International Joint Conference on Artificial Intelligence (IJCAI), 2015.}


% %% or include affiliations in footnotes:
% \author[mymainaddress,mysecondaryaddress]{Elsevier Inc}

% \author[mysecondaryaddress]{Global Customer Service\corref{mycorrespondingauthor}}
% \cortext[mycorrespondingauthor]{Corresponding author}

% \address[mymainaddress]{1600 John F Kennedy Boulevard, Philadelphia}
% \address[mysecondaryaddress]{360 Park Avenue South, New York}

\begin{abstract}
  Limited lookahead has been studied for decades in perfect-information games. We initiate a new direction via two simultaneous deviation points: generalization to imperfect-information games and a game-theoretic approach. We study how one should act when facing an opponent whose lookahead is limited. We study this for opponents that differ based on their lookahead depth, based on whether they, too, have imperfect information, and based on how they break ties.  We characterize the hardness of finding a Nash equilibrium or an optimal commitment strategy for either player, showing that in some of these variations the problem can be solved in polynomial time while in others it is PPAD-hard, NP-hard, or inapproximable. We proceed to design algorithms for computing optimal commitment strategies---for when the opponent breaks ties favorably, according to a fixed rule, or adversarially. We then experimentally investigate the impact of limited lookahead. The limited-lookahead player often obtains the value of the  game if she knows the expected values of nodes in the game tree for some equilibrium---but we prove this is not sufficient in general. Finally, we study the impact of noise in those estimates and different lookahead depths.
\end{abstract}
% % Keywords command
% \providecommand{\keyword}[1]
% {
%   \small	
%   \textbf{\textit{Keywords---}} #1
% }
% \begin{keyword}
% Game theory, equilibrium finding, limited lookahead, imperfect-information game, Nash equilibrium, Stackelberg equilibrium
% \end{keyword}

\section{Introduction}
Limited lookahead has been a central topic in AI game playing for decades. To
date, it has been studied in single-agent settings and perfect-information
games---specifically in well-known games such as chess, checkers, Go, etc., as
well as in random game tree
models~\cite{Berliner77:Search,Korf90:Real,Pearl81:Heuristic,Pearl83:Nature,Nau83:Pathology,Jansen90:Problematic,Nau10:When,Bouzy01:Computer,Ramanujan10:Adversarial,Ramanujan11:Trade}.
In this paper, we initiate the game-theoretic study of limited lookahead in
imperfect-information games. Such games are significantly more broadly
applicable to practical settings---for example auctions, negotiations, military
settings, security, cybersecurity, and medical settings---than
perfect-information games. \citet{Mirrokni12:Theoretical} conducted a
game-theoretic analysis of lookahead, but they consider only
perfect-information games, and the results are for four specific games rather
than broad classes of games. Instead, we analyze the questions for imperfect
information and for general games. Specifically, we study general-sum
extensive-form games. As is typical in the literature on limited lookahead in
perfect-information games, we derive our results for a two-agent setting. One
agent is a rational player (Player {\rational}) trying to optimally exploit a
limited-lookahead player (Player {\la}).

% Our results extend immediately to one rational player and more than one limited-lookahead player, as long as the latter all break ties according to the same scheme (statically, favorably, or adversarially---as described later in the paper). This is because such a group of limited-lookahead players can be treated as one from the perspective of our results. %, as described later).

The type of limited-lookahead player we introduce is quite natural and analogous to that in the literature on perfect-information games.  Specifically, we let the limited-lookahead player {\la} have a node evaluation function that places numerical values on all nodes in the game tree. Given a strategy for the rational player, at each information set at some depth $i$, Player {\la} picks an action that maximizes the expected value of the evaluation function at depth $i+k$, assuming optimal
%hypothetical
play between those levels.

% This model has applications in multiagent settings where a rational agent interacts with myopic or semi-myopic agents. For example, biological games where the goal is to steer evolution towards some desired outcome.
Our study is the game-theoretic, imperfect-information generalization of lookahead questions studied in the literature and therefore interesting in its own right.
We are motivated by three things: heuristic search algorithms, biological games, and security games.
In terms of heuristic search, we think that our model may be relevant to how one might incorporate opponent models into newer research on search in EFGs.
In biological games, the goal is to steer an evolution or adaptation process (which typically acts myopically with lookahead 1)~\cite{Sandholm12:Medical,Sandholm15:Medical,Kroer16:Sequential}. For example, if we wish to model a biological steering process such as in \citet{Kroer16:Sequential}, but with statistical guarantees over model uncertainty (such as in \cite{Chen12:Tractable}), then a limited lookahead model with information sets representing uncertainty and limited lookahead representing biological adaptation would allow capturing both those aspects.
For security settings, our results are most likely to be useful for settings where simple-minded agents are considered. For example, lookahead~1 agents seem like a reasonable model in sequential security settings, where adversaries might condition on their belief over possible states of the world, while not using sophisticated reasoning about the future steps of the game (e.g. fare-evasion games~\cite{Yin12:TRUSTS})).
Furthermore, investigating how well a rational player can exploit a limited-lookahead player lends insight into the limitations of using limited-lookahead algorithms in multiagent decision making.

We consider the problem of exploiting a limited-lookahead opponent under various assumptions about the opponent, mapping out the hardness of the problem under all these alternative assumptions. We consider three dimensions: whether the opponent has information sets, whether the opponent has lookahead $1$ or more, and whether the opponent breaks ties statically, adversarially, or favorably.
% The tie-breaking scheme employed may be critical in determining the
% optimal strategy to commit to. Any optimal strategy for some tie-breaking scheme can have arbitrarily large regret under another tie-breaking scheme.
%The appropriate tie-breaking assumption
%will necessarily depend on the application at hand. If robustness
%is desired, adversarial tie-breaking would probably be assumed.
If Player {\la} has no information sets, lookahead $1$, and breaks ties either adversarially or by a static scheme, we show that both a Nash equilibrium and an optimal strategy to commit to (that is, a Stackelberg strategy) can be found in polynomial time. Conversely, if any of these assumptions do not hold, we show that equilibrium finding is PPAD-hard and finding an optimal strategy to commit to is NP-hard.
This extends the study of exploiting limited-lookahead adversaries from the perfect-information setting. In the perfect-information setting,
\citet{korf1989generalized} studies generalized game trees where the players have different evaluation functions in the context of minimax search, and
\citet{Carmel96:Incorporating} study how opponent models can be incorporated into minimax search. In imperfect-information games, minimax search does not apply because the game tree does not decompose into subgames that can then be solved using information only from that subtree. Rather, strategies have to be ``balanced'' at all parts of the game tree holistically. For example, in poker, one cannot simply bet the good hands and call the bad hands: that would be too transparent and the opponent could easily exploit such a strategy.
\footnote{\citet{Frank98:Search} extend minimax-style opponent modeling to EFGs via Stackelberg equilibrium (although it is not described as such), but require the follower to have perfect information, and they use enumeration over leader strategies.}

We then design algorithms for finding an optimal strategy to commit to for the unlimited, rational player  {\rational}. We focus on this rather than equilibrium computation because the latter seems nonsensical in this setting: the limited-lookahead player determining a Nash equilibrium strategy would require her to reason about the whole game for the rational player's strategy, which rings contrary to the limited-lookahead assumption.
 %(although one could try to make these consistent by perhaps having the equilibrium computation occur offline and the lookahead online.  STILL WOULD NOT MAKE SENSE. ???
Furthermore, optimal strategies to commit to are desirable for applications such as biological games (because evolution is responding to what we as the ``steerer" are doing) and security games (where the defender typically gets to commit to a strategy).
Computing optimal strategies to commit to in standard rational settings has previously been studied in normal-form games \cite{Conitzer06:Computinga} and extensive-form games \cite{Letchford10:Computing}, the latter implying some complexity results for our setting as we will discuss.

For the case where the limited-lookahead player breaks ties in favor of Player {\rational}, or by some static scheme, we develop a mixed-integer program (MIP) that is a natural extension of the sequence-form linear program (LP) from the two-player zero-sum setting.

Then, we derive an algorithm for solving the setting where the limited-lookahead player breaks ties adversarially.  For a \emph{given} set of actions that are optimal for the limited-lookahead player, this ends up being a zero-sum game between the rational player and the tie-breaking rule. We then show how to embed this LP in a MIP that branches on which action set to make optimal for the limited-lookahead player.
% When ties are broken adversarially, we prove that Player {\rational} choosing some action set to entice Player {\la} to play induces a zero-sum game that can be solved by a linear program that is similar to the standard sequence-form linear program for solving two-player zero-sum extensive-form games \cite{Stengel96:Efficient}. Using this result, we develop a mixed-integer program (MIP) for finding the optimal strategy to commit to for Player {\rational}. We also develop MIPs for computing the optimal strategy to commit to both for favorable and static tie breaking.

We experimentally evaluate the usefulness of exploiting limited-lookahead opponents in two recreational games using our new algorithms. The limited-lookahead player often obtains the value of the  game if she knows the expected values of nodes in the game tree for some equilibrium---but we provide a counterexample that shows that this is not sufficient in general. We go on to study the impact of noise in those estimates, and different lookahead depths.
% We find that a lookahead pathology occurs naturally in the context of using Nash equilibrium as the node-evaluation function.

As in the literature on lookahead in perfect-information games, a potential weakness of our approach is that we require knowing the $h$ function (but make no other assumptions about what information $h$ encodes). In practice, this function may not be known. As in the perfect-information setting, this can lead to the rational exploiter being exploited. However, many practical settings do not have this problem. For example, biological design games~\cite{Sandholm12:Medical,Sandholm15:Medical} and fare-inspection games~\cite{Yin12:TRUSTS} involve myopic agents that would not be expected to design strategies that exploit the rational player's errors in beliefs about $h$.  If there are multiple limited-lookahead players, it seems even less likely that they could
% (optimally)
exploit the rational player in this way, as it may require coordination/cooperation.
%Generally, any time the limited-lookahead opponent is modeling a group of myopic agents that are not operating as a single unit, it would be unlikely that they would be able to exploit the strategies we devise.

In general, this paper can be taken as a prescriptive theory of how one should play against a limited-lookahead player, and how a limited-lookahead player should play, or as an investigation of how badly a best-responding limited-lookahead player can be exploited.

\subsection{Subsequent research on depth-limited solving in imperfect-information games}
Since the conference version of this paper, forms of search have been introduced for imperfect-information games. That work is quite different than that in the present paper: those newer papers attempt to approximate Nash equilibrium rather than working explicitly as, or against, a depth-limited player. Here we nevertheless briefly discuss those search approaches due to their success.

In one strand, \emph{blueprint strategies} for the players are computed in an abstraction of the entire game. Then, at each step as the game progresses, a finer-grained abstraction of the remaining game is solved (because that is computationally feasible as there is less and less game tree left to consider)~\cite{Brown17:Safe}. The key is to give the opponent the virtual choice of playing into the original blueprint game or the more refined abstraction of the remaining game. If we can compute a strategy for ourselves for the refined abstraction so that the opponent does not have incentive to play into the refined abstraction for any private type she might have, we have guaranteed that this nested endgame solving approach cannot make the strategies worse (that is, more exploitable) than the blueprint strategies. So, this approach is safe in that sense. More importantly, in practice it leads to strategies that are significantly stronger than the blueprint strategies. An additional improvement is to take into account that sometimes we can determine a lower bound on the gifts that the opponent has given us so far in his moves down the game tree through mistakes. That amount can be safely given back to the opponent. This enlarges the strategy space that can safely be optimized over in the refined abstraction of the remaining game, thereby leading to even stronger strategies in practice. These techniques were a key part of \emph{Libratus}, the first superhuman AI for two-player no-limit Texas hold'em poker~\citep{Brown18:Depth}.

Such search approaches can also be made to terminate before a leaf of the game tree. There are currently two approaches for doing so.

In one~\citep{Moravcik17:DeepStack}, large numbers of randomly generated situations of later possible situations (where the situation includes not only publicly observable aspects of state but also the players' belief distributions) are solved in advance. Then, when search is conducted, the search is terminated when it hits the depth of such previously-solved situations. Deep learning can be used to generalize to situations that were not in the set of solved situations.
If one were to incorporate opponent modeling into such an approach, then our results show that it potentially opens one up to exploitation.

A more recent approach is to, at each node at the depth limit of the search, allow each agent to select from a set of pre-computed \emph{continuation strategies} for the remaining game~\cite{Brown18:Depth}. These choices, in effect, prevent the searcher from making overly optimistic choices in the tree between the current state and the depth limit of the search. As enough continuation strategies are generated, the approach provably leads to Nash equilibrium in two-player zero-sum games. In practice it leads to very strong strategies already with a very small number of pre-computed continuation strategies and works beyond two-player zero-sum games. This approach was the key technique in \emph{Pluribus}, the first superhuman AI for multi-player no-limit Texas hold'em poker~\cite{Brown19:Superhuman}.

\section{Extensive-form games}

We start by defining the class of games that the players will play, without reference to limited lookahead. The class is general and standard.

An \emph{extensive-form game} $\Gamma$ is a tuple $\langle N,A,S,Z,\mathcal{H},\sigma_0,u,\mathcal{I}\rangle$. $N$ is the set of players. $A$ is the set of all actions in the game. $S$ is a set of nodes corresponding to sequences of actions. They describe a tree with root node $s^r\in S$. At each node $s$, it is the turn of some Player $i$ to move. Player $i$ chooses among actions $A_s$, and each branch at $s$ denotes a different choice in $A_s$. Let $\child{s}{a}$ be the node transitioned to by taking action $a\in A_s$ at node $s$. The set of all nodes where Player $i$ is active is called $S_i$.
$Z\subset S$ is the set of leaf nodes, where $u_i(z)$ is the utility to Player $i$ of node $z$. We assume, without loss of generality, that all utilities are non-negative. $Z_s$ is the subset of leaf nodes reachable from a node $s$. $\mathcal{H}_i\subseteq \mathcal{H}$ is the set of heights in the game tree where Player $i$ acts. $\mathcal{H}_0$ is the set of heights where Nature acts. $\sigma_0$ specifies the probability distribution for Nature, with $\sigma_0(s,a)$ denoting the probability of Nature choosing outcome $a$ at node $s$.

Imperfect information is represented in the game model using information sets.
$\mathcal{I}_i\subseteq\mathcal{I}$ is the set of information sets where Player $i$ acts. $\mathcal{I}_i$ partitions $S_i$. For nodes $s_1,s_2 \in I,I\in\mathcal{I}_i$, Player $i$ cannot distinguish among them, and $A_{s_1}=A_{s_2}$. %We let $X(s)$ denote the set of information set and action pairs $I,a$ in the sequence leading to a node $s$, including Nature. We let $X_{-i}(s),X_i(s)\subseteq X(s)$ be the subset of this sequence  such that actions by the subscripted player(s) are excluded or exclusively chosen.

We denote by $\sigma_i$ a {\em behavioral strategy} for Player $i$. For each information set $I\in \mathcal{I}_i$, it assigns a probability distribution over $A_I$, the actions at the information set. $\sigma_i(I,a)$ is the probability of playing action $a$.
% We let $\sigma_i(s,a)=\sigma(I,a)$ for all $s\in I$.
A {\em strategy profile} $\sigma=(\sigma_0,\ldots,\sigma_n)$ consists of a behavioral strategy for each player. We will often use $\sigma(I,a)$ to mean $\sigma_i(I,a)$, since the information set specifies which Player $i$ is active.
As described above, randomness external to the players is captured by the Nature outcomes $\sigma_0$. Using this notation allows us to treat Nature as a player when convenient, although Nature selects actions according to fixed probabilities.
%(rather than based on selecting its strategy based on the players' strategies).
%
%We let $\sigma_{I\rightarrow a}$ denote the strategy profile obtained from $\sigma$ by having Player $i$ deviate to taking action $a$ at $I\in \mathcal{I}_i$.

Let the probability of going from node $s$ to node $\hat{s}$ under strategy profile $\sigma$ be
  $\pi^\sigma(s, \hat s)=\Pi_{\tuple{\bar s, \bar a} \in X_{s, \hat s}} \sigma (\bar s, \bar a)$
where $X(s, \hat s)$ is the set of pairs of nodes and actions on the path from $s$ to $\hat s$.
We let the probability of reaching node $s$ be $\pi^\sigma (s)=\pi^\sigma(s^r, s)$, the probability of going from the root node to $s$. Let $\pi^\sigma (I)=\sum_{s\in I} \pi^\sigma (s)$ be the probability of reaching any node in $I$. $\pi^{\sigma}_i(I)=\pi^{\sigma}_i(s)\forall s\in I$ due to perfect recall. For probabilities over Nature, $\pi^\sigma_0=\pi^{\bar{\sigma}}_0$ for all $\sigma,\bar\sigma$, so we can ignore the strategy profile superscript and write $\pi_0$.
Finally, for all behavioral strategies, the subscript $-i$ refers to the same definition, excluding Player $i$. For example, $\pi_{-i}^\sigma(s)$ denotes the probability of reaching $s$ over the actions of the players other than $i$, that is, if $i$ played to reach $s$ with probability~$1$.

%The same elements are defined for $M'$, except each one is denoted by
%a prime superscript. For example, the node space for $M'$ is denoted
%by $S'$. The value for Player $i$ at leaf node $z^{\prime} \in Z^{\prime}$ in $M^{\prime}$ is denoted $W_i(z^{\prime})$.

%We let the height $k$ of a node $s$ be the length of the path from the node to the leaf level. Thus, the height of a leaf node is $0$. We assume that all leaves are at the same depth in the tree, and at each height only one player has active information sets. This is without loss of generality since for any game that does not satisfy this, we can modify it by inserting singleton information sets with only one action available to stretch out the game until the property is satisfied; this keeps the game strategically equivalent. Let $\mathcal{H}_i$ be the set of heights where player $i$ is active, and let $\mathcal{H}_i^l$ be $\{k:k\in \mathcal{H}_i, k\leq l\}$. Let $S_k$ and $\mathcal{I}_k$ be the sets of nodes and information sets at height $k$, respectively.

%For information set $I$ and action $a\in A_I$ at level $k\in \mathcal{H}_i$, we let $\mathcal{D}_{I}^{a}$ be the set of information sets at the next level in $\mathcal{H}_i$ reachable from $I$ when taking action $a$. Similarly, we let $\mathcal{D}_I^l$ be the set of descendant information sets at height $l\leq k$, where $\mathcal{D}_I^k=\{I\}$.

\section{Model of limited lookahead}

We now describe our model of limited lookahead, % for extensive-form games.
which we consider to be very intuitive.

We use the term optimal \emph{hypothetical} play to refer to the way the limited-lookahead agent thinks she will play when looking ahead from a given information set.  In actual play part way down that plan, she may change her mind because she will then be able to see to a deeper level of the game tree (given that her lookahead depth is still the same and she will be at a deeper part of the tree).

\begin{sloppypar}
Let $k$ be the lookahead of Player {\la}, and $S_{I,a}^k$ the nodes at the lookahead depth $k$ below information set $I$ that are reachable (through some path) by action $a$.
% As in prior work on lookahead in the perfect-information game setting, Player {\la} has a node-evaluation function $h:S\rightarrow \mathbb{R}$ that assigns a heuristic numerical value to each node in the game tree.
As in prior work in the perfect-information game setting, Player {\la} has a node-evaluation function $h:S\rightarrow \mathbb{R}$ that assigns a heuristic numerical value to each node in the game tree.
\end{sloppypar}
% Change

Given a strategy $\sigma_\rational$ for the other player and fixed action probabilities for Nature,  Player {\la} chooses, at any given information set $I\in \mathcal{I}_{\la}$ at depth $i$, a (possibly mixed) strategy whose support is contained in the set of actions that maximize the expected value of the heuristic function at depth $i+k$, assuming optimal hypothetical play by her ($\max_{\sigma_l}$ in the formula below).  We will denote this set by 
\vspace{-2mm}
\begin{align*}
A_I^*=
  \{a:a\in \arg\max_{a\in A_I, \sigma_{\la}} \sum_{s\in I}\frac{\pi^{\sigma_{-\la}(s)}}{\pi^{\sigma_{-\la}}(I)}\sum_{s'\in S_{I,a}^{k}} \pi^{\sigma}(\child{s}{a},s') h(s') \},
\vspace{-2mm}
\end{align*}
where $\sigma=\{\sigma_{\la},\sigma_{\rational}\}$.
 %and $S^i$ is the set of nodes at depth $i$.
%
Here moves by Nature are also counted toward the depth of the lookahead of the limited-lookahead player, and when looking through such nodes, that player takes an expectation over Nature's moves at that node.

The model is flexible as to how the rational player chooses $\sigma_\rational$ and how the limited-lookahead player chooses a (possibly mixed) strategy with supports within the sets $A_I^*$.  For one, we can have these choices be made for both players simultaneously according to the Nash equilibrium solution concept, so neither player wants to change her choices given that the other does not change.  As another example, we can ask how the players should make those choices if one of the players gets to make, and commit to, all her choices before the other. This begets multiple settings based on which player gets to commit first and how ties are broken.  We will study all of the above variants.  Other solution concepts and refinements could also be used.

\begin{figure}[!h]
  %\vspace{-15mm}
  \centering
    \includegraphics[scale=0.51]{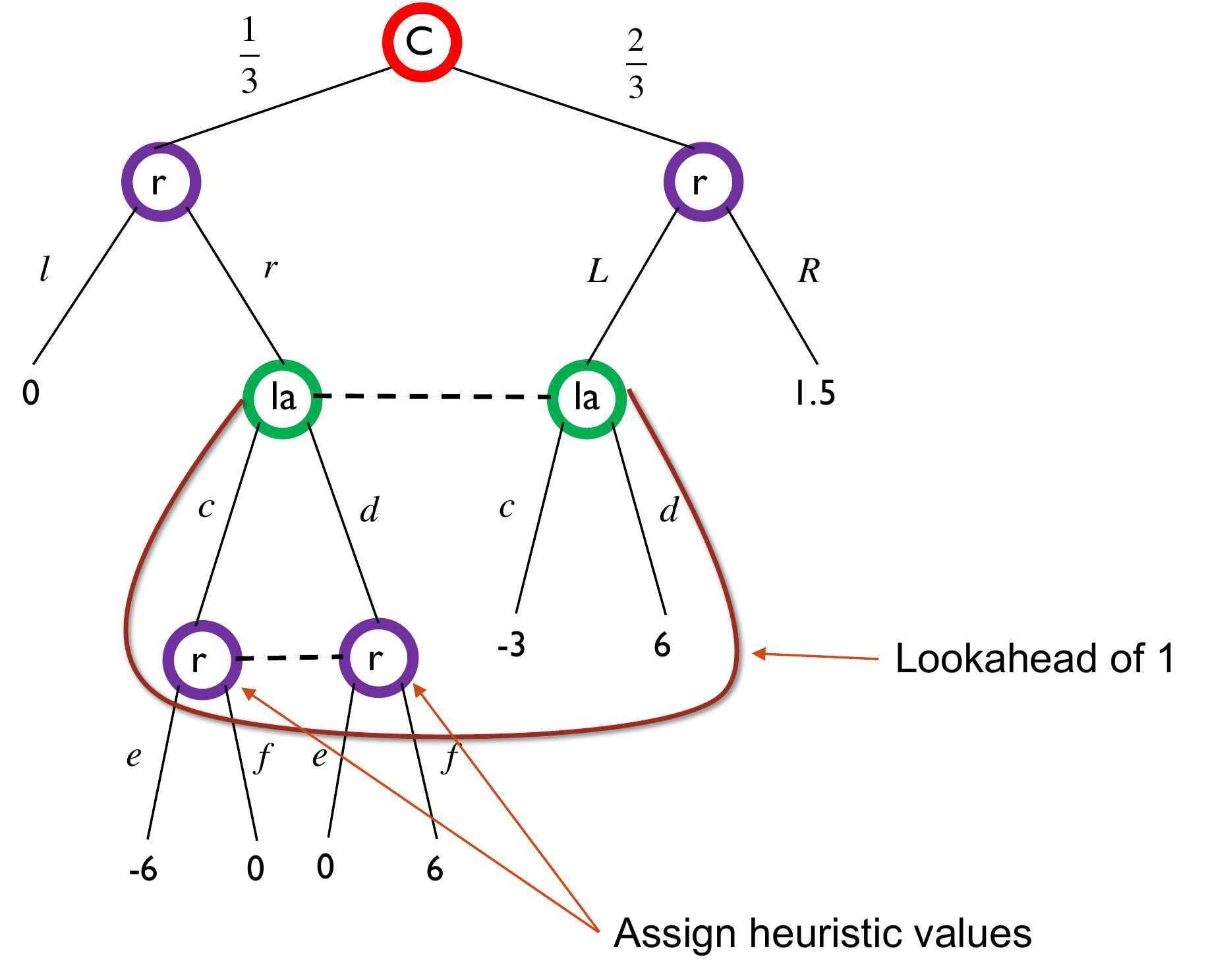}
    %\vspace{-3mm}
  \caption{An example of a limited-lookahead game tree.}
  \label{fig:lookahead_tree_example}
%\vspace{-4mm}
\end{figure}

An example is given in Figure~\ref{fig:lookahead_tree_example}. In this game the first player is a rational player, while the second player is a limited-lookahead player. Player {\la} has only a single information set, and the lookahead boundary for that information set is denoted by a curbed line. When making decisions at their information set, Player {\la} will maximize the expected value of their heuristic function over the nodes within the decision boundary. The two leaf nodes within the boundary are assigned the correct value, whereas the two nodes belonging to P1 would be assigned values according to the node-evaluation function.

%Throughout this paper, we will assume that Player {\la} has limited lookahead,
%while Player {\rational} is a standard rational player. We also assume perfect
%recall for Player {\rational} (or both?).

\section{Complexity}
In this section we analyze the complexity of finding strategies according to these solution concepts.

\subsection{Nash equilibrium}
\begin{proposition}
 Finding a Nash equilibrium when Player {\la} either has information sets containing more than one node, or has lookahead at least $2$, is PPAD-hard.
\end{proposition}
 This is because finding a Nash equilibrium in a 2-player general-sum normal-form game is PPAD-hard~\cite{Chen09:Settling}, and any such game can be converted to a depth $2$ extensive-form game (where the second player does not know what the first player moved), where the general-sum payoffs are the evaluation function values.

 \begin{proposition}
 If the limited-lookahead player only has singleton information sets and lookahead 1, an optimal strategy can be trivially computed in polynomial time   in the size of the game tree.
 \end{proposition}
 For each of her information sets, we simply have to pick an action that has highest immediate heuristic value.  To get a Nash equilibrium, what remains to be done is to compute a best response for the rational player, which can also be easily done in polynomial time~\cite{Johanson11:Accelerating}.

\subsection{Commitment strategies}
\begin{figure}[h]
  %\vspace{-2mm}
  \centering
        \scalebox{1.25}{
        \input{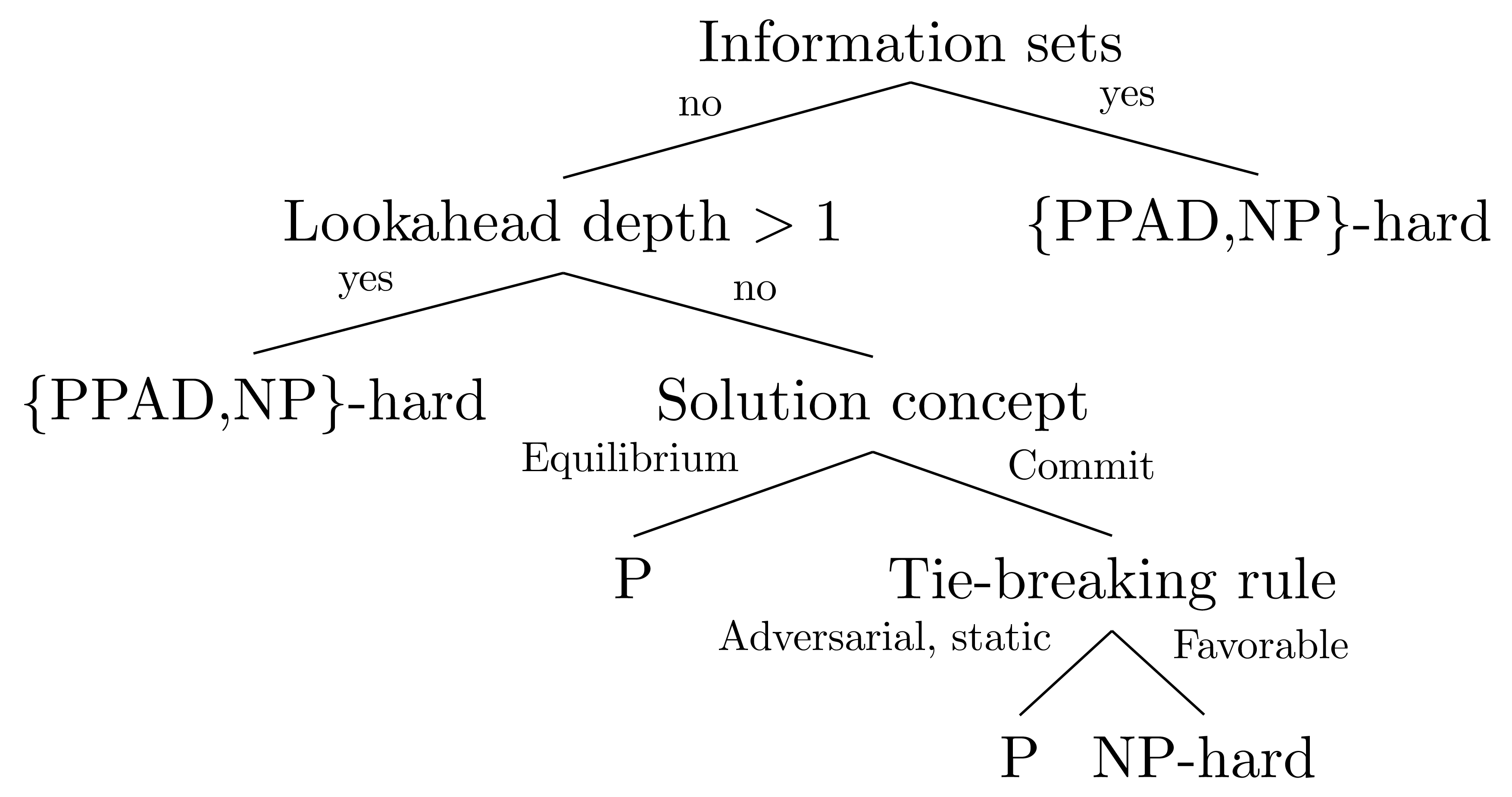}
    }
  %\vspace{-3mm}
  \caption{Our complexity results. \{PPAD,NP\}-hard indicates that finding a Nash equilibrium is PPAD-hard and finding an optimal strategy to commit to is NP-hard. P indicates polynomial time.
  }
  \label{fig:complexity-overview}
  %\vspace{-2mm}
\end{figure}

Next we study the complexity of finding commitment strategies. The complexity depends on whether the game has imperfect information (information sets that include more than one node) for the limited-lookahead player, how far that player can look ahead, and how she breaks ties in her action selection.
Figure~\ref{fig:complexity-overview} shows an overview of the different results that we show.

\vspace{1mm}\noindent\textbf{No information sets, lookahead 1, static tie-breaking}
\begin{proposition}
  If the limited-lookahead player only has singleton information sets and lookahead 1, an optimal strategy can be computed in polynomial time.
\end{proposition}
We can use the same approach as for the Nash equilibrium case, except that the specific choice among the actions with highest immediate value is dictated by the tie-breaking rule.
%in the size of the game tree for the limited-lookahead player (without even knowing the other player's strategy $\sigma_\rational$) because for each of her information sets, we simply have to pick the action that has highest immediate heuristic value.  To get a Nash equilibrium, what remains to be done is to compute a best response for the rational player, which can also be easily done in polynomial time.
% We start out by investigating the case where the player with
% limited lookahead has no information sets. This setting could be
% important for modeling limited-lookahead agents such as diseases and other
% biological applications, where information sets may not be present.
%For this case, a deterministic strategy
%for Player {\la} can inductively be found by considering each node separately. For each node, the tie-breaking scheme dictates some action from $A_{s}^{*}$, since the strategy of Player {\rational} has no effect on the heuristic utility.
%This is because the choice is based on a static tie-breaking scheme,
%and so our choice of strategy has no effect on the chosen strategy.
With this strategy in hand, finding a utility-maximizing strategy
for Player {\rational} again consists of computing a best response.

\vspace{1mm}\noindent\textbf{No information sets, lookahead 1, adversarial tie-breaking}
When ties are broken adversarially, the choice of response
depends on the choice of strategy for the rational player. The set of optimal actions $A_{s}^{*}$ for any node $s\in S_{{\la}}$ can be precomputed, since Player {\rational} does not affect which actions are optimal.
Player {\la} will then choose actions from these sets to minimize the utility of Player {\rational}. We can view
the restriction to a subset of actions as a new game, where Player {\la} is a rational player in a zero-sum game.
 % trying to minimize the utility of the other player.
An optimal strategy for Player {\rational} to commit to is
then a Nash equilibrium in this smaller game. This is solvable in
polynomial time by an LP that is linear in the size of the game
tree \citep{Stengel96:Efficient}, and algorithms have been developed
for scaling to large games \citep{Hoda10:Smoothing,Zinkevich07:Regret,Lanctot09:Monte}.

\vspace{1mm}\noindent\textbf{No information sets, lookahead 1, favorable tie-breaking}
In this case,
Player {\la} picks the action from $A_{s}^{*}$ that maximizes the utility of Player {\rational}. Perhaps surprisingly,
computing the optimal solution in this case is harder than when facing
an adversarial opponent. All our hardness proofs are by reduction from $3$SAT.
\begin{definition}
  A $3$SAT instance consists of a tuple $(V,C)$. $V$ is a set of $n$ Boolean variables, and $C$ is a set of $m$ clauses of the form $\left(l_1 \lor l_2 \lor l_3\right)$ where each $l_i$ represents a literal requiring some variable to be true or false.
\end{definition}

We will also use a variant of 3SAT: MAXSAT. In MAXSAT the goal is to find an assignment maximizing the number of satisfied clauses.
In the case where each clause has 3 literals, it is known that MAXSAT is NP-hard, and unless P=NP, there is no approximation algorithm with a performance ratio better than $\frac{7}{8}$~\citep{Haastad01:Some}.

\begin{theorem}
Computing a utility-maximizing strategy for the rational player to
commit to is inapproximable to a factor better than $\frac{7}{8}$
if the limited-lookahead player breaks ties
in favor of the rational player, unless P=NP. \label{the:break-ties-in-favor}
\end{theorem}
\begin{proof}
  We reduce from MAXSAT. A picture illustrating our reduction is given in Figure~\ref{fig:3sat-break-ties-favorably}, and a description is given below.

  Let the root node be a chance node. It chooses with equal probability
between $\left|C\right|$ child nodes, each representing a clause.
Each such descendant clause node is a singleton information set belonging
to Player {\la}.
Each clause node has three actions, representing the three literals in the clause. Each such action leads to a node representing that literal. Player {\la} gets the same value from each action and is therefore indifferent. Player {\rational} acts at each
literal node, with all literal nodes representing the same variable being in an information set together. Thus, Player {\rational} has an information set for each variable. At each variable information set, there is a true and false action. For a given literal node in some variable information set, the true action leads a payoff of $1$ if the literal requires the variable to be true, and $0$ otherwise. Similarly, the false action leads to a payoff of $1$ if the literal requires the variable to be false, and $0$ otherwise.
%Each variable node has two actions, true and false,
%with payoff $1$ to Player {\rational}, and payoff $0$ or $1$ to Player {\la},
%depending on whether the literal is in the clause or not, respectively.
%Player {\rational} has an information set for each variable, containing all
%variable nodes representing said variable.

The decision problem is then: does there exist a strategy for Player {\rational} with expected payoff $\frac{k}{n}$? If there is a MAXSAT assignment with $k$ satisfied clauses then we can take that assignment and use it as the strategy for the follower. For the leader, we pick one of the satisfied literals at every clause information set, and an arbitrary action at unsatisfied clauses. This clearly achieves utility $\frac{k}{n}$.
Conversely, say there is a strategy profile $\sigma$ achieving utility $\frac{k}{n}$. First, we may assume that the follower strategy is a pure strategy: if not then there exists some pure strategy which is still a best response, with at least as high utility to the leader. Now say the leader strategy is a mixed strategy, and let $c_i$ be some clause with nonzero probability on variables $v,v'$. In that case either $v$ or $v'$ achieves weakly greater utility than the other, say $v$ does. Now set the probability of $v$ to $1$ at $c_i$. Since the follower gets the same utility everywhere their best response set does not change, and we have weakly increased the utility of the leader. Thus we may also assume that the leader strategy is pure. This yields a pair of pure strategies achieving utility $\frac{k}{n}$. We can now construct a true/false assignment such that at least $k$ clauses must be satisfied, corresponding to the ones achieving nonzero utility in the game.
\end{proof}
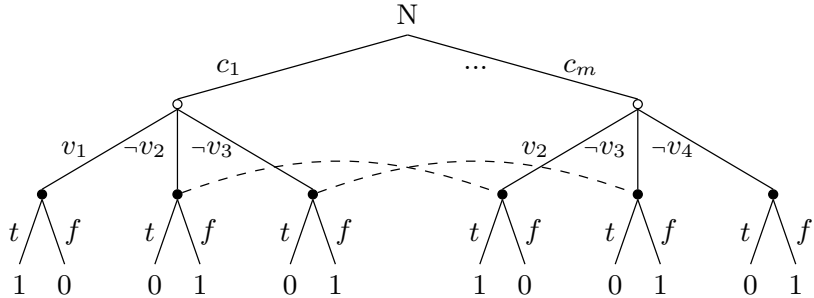
\begin{figure}[!h]
  \begin{center}
        \scalebox{1.20}{
        \input{figures/3sat-break-ties-favorably}
    }
  \end{center}
  % \includegraphics[scale=0.3]{figures/tight-action-abstraction}
%  %\vspace{-.15in}
  \caption{The game tree in our proof of Theorem~\ref{the:break-ties-in-favor}. Dashed lines denote information sets. %N denotes a Nature node, hollow nodes denote Player {\la} nodes and solid nodes denote Player {\rational} nodes.
  }
  \label{fig:3sat-break-ties-favorably}
\end{figure}

\begin{figure}[!h]
  \begin{center}
        \scalebox{1.20}{
        \input{figures/3sat-look-ahead-2}
    }
  \end{center}
  % \includegraphics[scale=0.3]{figures/tight-action-abstraction}
%  %\vspace{-.15in}
  \caption{The clause modification in our proof of Theorem~\ref{the:np-hard-lookahead-2}.
  }
  \label{fig:3sat-look-ahead-2}
\end{figure}
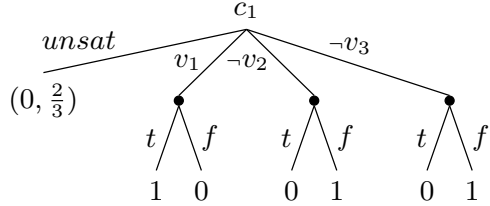

\vspace{1mm}\noindent\textbf{No information sets, lookahead $>$ 1, any tie-breaking}
It is NP-hard to compute an optimal
strategy to commit to in extensive-form games when both players are
rational~\cite{Letchford10:Computing}. That was proven by reducing from knapsack to a 2-player perfect-information game of depth $4$.
% This immediately gives us two results:
% (1) finding an optimal strategy for Player {\rational} to commit to is NP-hard if Player {\la} has lookahead at least $4$, and (2) computing an optimal strategy to commit to for Player {\la} is NP-hard even with lookahead $1$.
Their result implies NP-hardness of computing a strategy to commit
to for the rational player, if our {\la} player has lookahead of
at least $4$. We tighten this to lookahead $2$:
\begin{theorem}
Computing a utility-maximizing strategy for the rational player to
commit to is NP-hard if the limited lookahead player has lookahead
at least $2$, no matter how they break ties. \label{the:np-hard-lookahead-2}
\end{theorem}
\begin{proof}
  We reduce from $3$SAT. We use the same reduction as for Theorem~\ref{the:break-ties-in-favor}, except that at each clause node, we also add an ``unsatisfied'' action that leads directly to a leaf node with payoff $0$ for Player {\rational} and payoff $\frac{2}{3}$ for Player {\la}.

  For all leaf nodes under a variable node, we set the payoff to $1$ for Player {\rational}, and $0$ or $1$ for Player {\la}, for when the leaf represents the ancestor clause being unsatisfied or satisfied by the literal, respectively. The modifications are shown for a single clause in Figure~\ref{fig:3sat-look-ahead-2}.
% Let the root node be a chance node. It chooses with equal probability
% between $\left|C\right|$ child nodes, each representing a clause.
% Each such descendant clause node is a singleton information set belonging
% to Player {\la}. It has an ``unsatisfied'' action that leads directly
% to a leaf node with payoff $(0,1)$ for Player {\rational} and Player {\la}, respectively.
% It also has an action for each literal in the clause, leading to a
% variable node corresponding to the literal. Player {\rational} acts at each
% variable node. Each variable node has two actions, true and false,
% with payoff $1$ to Player {\rational}, and payoff $0$ or $1$ to Player {\la},
% depending on whether the literal is in the clause or not, respectively.
% Player {\rational} has an information set for each variable, containing all
% variable nodes representing said variable.

The question is whether Player {\rational} can compute a strategy such that
Player {\la} selects a literal action for each clause. For a given variable, choosing a probability strictly between
$0,\frac{2}{3}$ for the two actions leads to zero utility gain, since Player
{\la} will then always prefer the unsatisfied actions over any literal
belonging to the variable. Thus we can assume that Player {\rational} plays
a pure strategy, since at most one action can have its probability set high enough to yield utility gain. Thus the tie-breaking rule does not matter, since ties cannot occur for the {\la} player when the rational player plays a pure strategy.
Now, for each clause, Player {\la} will only choose
a literal action if that variable is set to the correct value to satisfy
the clause. Thus, if Player {\rational} can compute a strategy that gives
expected utility $1$, each clause node must have at least one variable with a satisfying assignment.
\end{proof}

\vspace{1mm}\noindent\textbf{Limited-lookahead player has information sets, lookahead $1$ and any tie-breaking rule}
When the limited lookahead player has information sets, we show that
computing a strategy to commit to is NP-hard:
\begin{theorem}
Computing a utility-maximizing strategy for the rational player to
commit to is NP-hard if the limited lookahead player has information
sets of at least size $6$, no matter the tie-breaking rule.
\label{the:np-hard-information-sets}
\end{theorem}
\begin{proof}
We reduce from $3$SAT. Let the root node be a chance node. It chooses
with equal probability between all variable and clause pairs $v,c$ such that $v\in c$. Player {\rational} acts at each child node,
being able to distinguish only which variable was chosen.
%Each information set thus has $\left|C\right|$ nodes.
For each information set, Player
{\rational} can choose between a true and a false action, representing setting
the associated variable to true or false, respectively. At the next
level where Player {\la} is active. The information sets at the level are
constructed as follows.
%If $v\notin c$, each node, whether Player
%$1$ picked true or false, is a singleton information set, with a
%single action leading to payoff $1$ for both players.
For each $c\in C$
an information set is constructed, containing all nodes representing
Player {\rational} choosing both true and false for each $v\in c$. For each
information set representing some clause $c$, Player {\la} has $4$
actions. First is an unsat action, leading to payoff $0$ for Player
{\rational} and payoff $\frac{2}{3}$ for Player {\la}, no matter which node in the
information set play has reached. Second, an action for each variable
$v\in c$ leading to payoff $1$ for Player {\rational}, and payoff $3$
to Player {\la} if play reached a node representing $v$ with true
or false chosen such that it satisfies $c$, and payoff $0$ for all
other nodes in the information set.

We claim that there is a satisfying assignment if and only if Player
{\rational} can commit to a strategy with expected payoff $1$. Let $\phi:V\rightarrow\{true,false\}$
be a satisfying assignment to $V,C$. Let Player {\rational} deterministically
pick actions at each variable information set according to $\phi$.
If play reaches a singleton node, Player {\la} has only one action
available, guarateeing payoff $1$. If play reaches some information
set representing a clause $c$, Player {\la} has expected payoff of
$3\cdot\frac{1}{3}$ when picking any action representing a satisfied
literal $l\in c$, as the conditional probability of being at a node
satisfying the clause is at least $\frac{1}{3}$, and Player {\rational} chooses the
satisfying action with probability $1$. % Since Player {\la} breaks
% ties such that unsatisfied actions are least preferred, she will pick
% an action representing a variable for each information set, yielding
% payoff $1$ to Player {\rational}.
This covers all possible outcomes, giving
an expected payoff of $1$ to Player {\rational}.

Given some strategy for Player {\rational} that gives payoff $1$ in expectation,
we show how to construct a satisfying assignment to $V,C$. For a strategy to
have payoff $1$, Player {\la} must be choosing variable actions at each
information set for some clause $c$. This is the case if and only if Player
{\rational} selects the satisfying truth value with probability at least
$\frac{2}{3}$ for some $v\in c$, since the expected payoff of taking a variable
action is otherwise strictly smaller than the unsatisfied action. This leads
directly to a satisfying assignment, by choosing the corresponding value
assignment for each action that is selected with probability $\frac{2}{3}$, and choosing
an arbitrary assignment for every other variable. This works no matter the tie-breaking rule, since Player {\rational} can always increase the probability to $1$ without changing their payoff.
\end{proof}
\begin{figure}[!h]
  \begin{center}
        \scalebox{1.20}{
        \input{figures/3sat-information-sets}
    }
  \end{center}
  \caption{The game tree for our proof of Theorem~\ref{the:np-hard-information-sets}.
  }
  \label{fig:3sat-information-sets}
\end{figure}
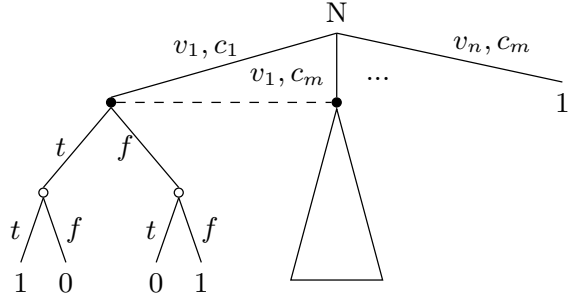

\section{Algorithms} \label{sec:algorithms}

We showed how to compute
an optimal strategy to commit to in polynomial time when the limited-lookahead player has no information sets, lookahead $1$,
and ties are broken either by a static scheme or adversarially. We
then showed hardness for all other cases.
In this section we will develop worst-case exponential-time algorithms
for solving the hard commitment-strategy cases.
%Later we will show that they scale to medium size game, in spite of the worst-case guarantees.
%
Here we focus on commitment strategies rather than the hard Nash equilibrium problem classes because Player {\la} playing a Nash equilibrium strategy would require the limited-lookahead Player {\la} to reason about the whole game for the opponent's strategy, which rings contrary to the limited-lookahead assumption.
%(although one could try to make these consistent by perhaps having the equilibrium computation occur offline and the lookahead onlien.  STILL WOULD NOT MAKE SENSE. ???
Further, optimal strategies to commit to are desirable for applications such as biological games (because evolution is responding to what we as the ``steerer" are doing) and security games (where the defender typically gets to commit to a strategy).

\subsection{Favorable tie breaking}

We start with the case where the limited-lookahead player breaks
ties in the rational player's favor. We use the idea of the sequence form~\cite{Romanovskii62:Reduction,Koller92:Complexity,Koller96:Efficient}, where a variable is
introduced for each {\em sequence} (information set-action pair) of actions a given player can take.
The insight is that in perfect-recall games, a given action
at some information set for Player $i$ is reached by a unique sequence of actions of Player $i$. This is exploited to represent the probability $\pi^{\sigma}_i(I)\sigma(I,a)$ of
a given action $a\in A_I$ being realized by a variable $x_{a}$. To
ensure that a valid set of realization probabilities is computed, the
constraint
$x_{a}=\sum_{a^{\prime}\in A_{I^{\prime}}}x_{a^{\prime}} $ is introduced for all information sets
$I^{\prime}$ and actions $a$ such that $a$ is the last action by
Player $i$ on the path to $I^{\prime}$.
A behavioral strategy is then obtained simply by normalizing by the
realization probability of the last action $a$: $\sigma(I^{\prime},a^{\prime})=\frac{x_{a^{\prime}}}{x_{a}}$.
With this formulation, duality is used to obtain a linear program
for computing Nash equilibria in zero-sum extensive-form games.

In our case, we cannot apply duality. Instead, we work directly on
the sequence form variables for both players. For Player {\rational}, we
introduce realization variables $x_{a}\in\left[0,1\right]$ for each
action $a$. For Player {\la}, we introduce Boolean realization variables
$y_{a}\in\left\{ 0,1\right\} $ for each action $a$, as there always
exists a pure strategy that maximizes utility, given a strategy for
the other player.
This is a key deviation from the traditional sequence form, where the variables are real valued.
%Equation \ref{equ:sequence-constraint} is used
%for both players, to ensure that a valid strategy profile is computed.

For any node $s$, we have $\pi_{1}(s)=x_{a},\pi_{2}(s)=y_{a^{\prime}}$
where actions $a,a^{\prime}$ are the last actions on the path to
$s$ for Player {\rational} and Player {\la}, respectively. Using this notation,
we introduce a variable $r_{z}$ representing the expected utility
from each leaf node $z$. The expected utility of a leaf node requires
computing the probability of it being reached $\pi_{0}(z)\cdot\pi_{1}(z)\cdot\pi_{2}(z)$,
which is a non-linear function.
However, since Player {\la} uses only probabilities $0$ and $1$,
we can separate this into two linear single-variable constraints
% for each leaf node $z$
%\vspace{-1.5mm}
\[
r_{z}\leq u_{{\rational}}(z) \pi_{0}(z) \pi_{\rational}(z) \quad \mbox{ and } \quad r_{z}\leq u_{\rational}(z)\pi_{\la}(z),
%\vspace{-1.5mm}
\]
where the first constraint is the reach probability times the utility when $\pi_{\la}(z)=1$, whereas the second forces $r_z$ to zero if the Boolean variable $\pi_{\la}(z)=0$.
The objective function is then simply $\sum_{z\in Z}r_{z}$.

Finally, we must ensure that the strategy chosen for Player {\la} maximizes
her utility according to the evaluation function at each information
set $I\in\mathcal{I}_{2}$. Let $S_{I,a}^{k}$ be the set of nodes
at depth $k$ below $I$, reachable from information set $I$ when taking action
$a$. Letting $\pi^\sigma (s)$ denote the probability of reaching $s$ under optimal hypothetical play, we introduce the following constraint for all $a,a^{\prime}\in A_{I}$:
\begin{align}
  %\vspace{-4mm}
\sum_{s\in S_{I,a}^{k}}\pi^\sigma(s) h(s)
  \geq&\sum_{s\in S_{I,a^{\prime}}^{k}}\pi^\sigma(s) h(s)-M(1-y_{a}) \label{equ:incentive-constraint-favorable}
  %\vspace{-4mm}
\end{align}
The constraint requires that the weighted sum over descendant node
evaluation function values is at least as high at $a$ as at any other
action $a^{\prime}$. The negative term ensures that the constraint
is active only if the action is chosen ($y_{a}=1$) by subtracting
a sufficiently large number $M$ otherwise.

The number of MIP matrix entries needed to implement this sparsely is
$O(\sum_{I\in \mathcal{I}_{\la}} \left|A_I\right| \cdot \max_{s \in S} \left| A_s \right|^{\min\{k,k^\prime\}})
$, where $k^\prime$ is the maximum depth of the subtrees rooted in $I$.
We present the details on the implementation and the proof of the MIP size in the proof of the similar case for Theorem~\ref{the:nash-equilibrium-induced-game}.
For many games, the lookahead depth $k$, maximum action set size, and number of information sets would all be much smaller than the size of the game tree $\left| S \right|$. For example, in the largest game that we investigate in the experimental section, the above expression, which is an upper bound, yields $448$ entries. The game tree has $199$ nodes.
The MIP is thus almost linear in the size of the game tree for many realistic games.

% \subsubsection{Specialization to the case of a rational follower}
% A special case of the MIP presented in the preceding section is the case where the follower is fully rational, i.e. does not suffer from limited lookahead.

% Concurrently with the conference version of this paper, \citet{Bosansky15:Sequence} used the same insight of using Boolean sequence-form variables for the follower in order to construct a linear-in-game-size MIP for computing Stackelberg equilibria in the standard setting.
% The result of \citet{Bosansky15:Sequence} and our MIP represent the first linear-in-game-size MIP for computing a Stackelberg equilibrium in an EFG without limited lookahead.

\subsection{Static tie-breaking}
When the limited-lookahead player breaks ties according to some static scheme $\succ$, we can use the same approach as for favorable tie breaking, except that Equation~\eqref{equ:incentive-constraint-favorable} has to be a strict inequality for any $a,a^\prime$ such that $a^\prime \succ a$.
This can be achieved in a MIP by subtracting sufficiently small $\epsilon$. In fact, most modern MIP solvers allow strict inequalities, and handle this implicitly without the user needing to find an appropriate $\epsilon$.
%To model this as a MIP, we keep the constraint as a weak inequality and subtract sufficiently small $\epsilon$ to get a strict inequality.
% to get the following for any such $a,a^\prime$:
% \begin{align}
%   &\sum_{s\in S_{sIa}^{k}}\pi_{0}(s)\cdot\pi_{1}(s)\cdot e(s) - \epsilon \nonumber\\
%   \geq&\sum_{s\in S_{I,a^{\prime}}^{k}}\pi_{0}(s)\cdot\pi_{1}(s)\cdot e(s)-M\cdot(1-y_{a}) \label{equ:incentive-constraint-static}
% \end{align}

\subsection{Adversarial tie breaking}
When the limited-lookahead player breaks ties adversarially, we wish to compute a strategy that
maximizes the worst-case best response by the limited-lookahead
player.
%In the previous two cases, only one action had to be chosen at each information set. For favorable tie-breaking, even if more than one action constitutes a best response, the maximization ensures that the correct one is chosen. %For the static tie-breaking scheme, this is enforced by the strict inequalities.
% In contrast, when ties are broken adversarially, we need to explicitly ensure that the minimizing action is chosen.% at each information set $I\in
%\mathcal{I}_2$.

For arguments sake, say that we were given $\mathcal{A}$, which is a fixed set of pairs, one for each information set $I$ of the limited-lookahead player, consisting of a set of optimal actions $A^*_I$ and one strategy for hypothetical play $\sigma^I_l$ at $I$. %(lookahead tree).
%choice of one or more actions at each information set for Player {\la}, including hypothetical play. %The choice of strategies  for the two players amount to a zero-sum game.
Formally,
$\mathcal{A} = \bigcup_{I\in \mathcal{I}_l} \langle A^*_I, \sigma^I_l \rangle$.
To make these actions optimal for Player {\la},
%the set of actions $\mathcal{A}$,
Player {\rational} must choose a strategy such that all actions in $\mathcal{A}$ are best responses according to the evaluation function of Player {\la}. Formally, for all $a,a^*\in \mathcal{A},a^\prime \notin \mathcal{A}$ (letting $\pi(s)$ denote probabilities induced by $\sigma_\la^I$ for the hypothetical play between $I,a$ and $s$):
\begin{align}
  %\vspace{-2.8mm}
  \sum_{s\in S_{I,a}^{k}}\pi(s)\cdot h(s)
  > \sum_{s\in S_{I,a^{\prime}}^{k}}\pi(s)\cdot h(s) \label{equ:dominated-actions} \\
  \sum_{s\in S_{I,a}^{k}}\pi(s)\cdot h(s)
  = \sum_{s\in S_{I,a^{*}}^{k}}\pi(s)\cdot h(s) \label{equ:incentivized-actions}
%\vspace{-1.1mm}
\end{align}
Player {\rational} has to choose a worst-case utility-maximizing strategy %(which is a Nash equilibrium strategy for two-player zero-sum games)
that satisfies Equations~\eqref{equ:dominated-actions} and~\eqref{equ:incentivized-actions}, and Player {\la} has to compute a (possibly mixed) strategy from $\mathcal{A}$ such that the utility of Player {\rational} is minimized.
We show that this problem can be solved by LP~\ref{equ:primal-lp-equilibrium}.
\begin{theorem}
  For some fixed choice of actions $\mathcal{A}$ to force Player {\la} to play, Nash equilibria of the induced game can be computed in polynomial time by a linear program that has size %TODO
$  O(\left|S\right|) +
O(\sum_{I\in \mathcal{I}_{\la}} \left|A_I\right| \cdot \max_{s \in S} \left| A_s \right|^{\min\{k,k^\prime\}})$.
  \label{the:nash-equilibrium-induced-game}
\end{theorem}
To prove this theorem, we first design a series of linear programs for computing best responses for the two players. We will then use duality to prove the theorem statement.

In the following, it will be convenient to change to matrix-vector notation. Our notation will be analogous to that of \citet{Stengel96:Efficient}, with some extensions. Let $A=-B$ be matrices describing the utility function for Player {\rational} and the maximization problem of Player~{\la} over $\mathcal{A}$, respectively. Rows are indexed by Player {\rational} sequences, and columns by Player {\la} sequences. For sequence form vectors $x,y$, the objectives to be maximized for the players are then $xAy,xBy$. Matrices $E,F$ are used to describe the sequence form constraints for Player {\rational} and {\la}, respectively. Rows correspond to information sets, and columns correspond to sequences. %For each information set $I$ and action $a\in A_I$, $F_{I,a}=1$, and for the parent sequence $a^\prime$, $F_{I,a^\prime} = -1$. $E$ is designed analogously.
Letting $e,f$ be standard unit vectors of length $\left| \mathcal{I}_{\rational}\right|, \left|\mathcal{I}_{\la} \right|$, respectively, the constraints $Ex=e,Fy=f$ describe the sequence form constraint for the respective players.
% Let $\mathcal{A}_I$ be the chosen set of actions for information set $I$.
Given a strategy $x$ for Player {\rational} satisfying Equations~\eqref{equ:dominated-actions} and~\eqref{equ:incentivized-actions} for some $\mathcal{A}$, the optimization problem for Player {\la} becomes choosing a vector of $y^\prime$ representing probabilities for all sequences in $\mathcal{A}$ that minimize the utility of Player {\rational}.
Letting a prime superscript denote the restriction of each matrix and vector to sequences in $\mathcal{A}$, this gives the following primal~(\ref{equ:primal-lp-best-response-p2}) and dual~(\ref{equ:dual-lp-best-response-p2}) LPs:
% \noindent\begin{minipage}{.5\linewidth}
% \begin{equation}
% \begin{aligned}
%   \max_{y^\prime} \quad &  (x^T B^\prime) y^\prime \\
%   & F^\prime y^\prime = f^\prime \\
% &  y\geq 0
% \end{aligned} \label{equ:primal-lp-best-response-p2}
% \end{equation}
% \end{minipage}%
% \begin{minipage}{.5\linewidth}
% \begin{equation}
% \begin{aligned}
%   \min_{q^\prime} \quad & q^{\prime T} f^\prime \\
%   & q^{\prime T} F^\prime \geq x^T B^\prime
% \end{aligned} \label{equ:dual-lp-best-response-p2}
% \end{equation}
% \end{minipage}
\vspace*{-8mm}
\begin{multicols}{2}
\begin{equation}
\begin{aligned}
  \max_{y^\prime} &\quad  (x^T B^\prime) y^\prime \\
  & F^\prime y^\prime = f^\prime \\
&  y\geq 0
\end{aligned} \label{equ:primal-lp-best-response-p2}
\end{equation}\break%\hspace{-20cm}
\begin{equation}
\begin{aligned}
  \min_{q^\prime}  & \quad q^{\prime T} f^\prime \\
  & q^{\prime T} F^\prime \geq x^T B^\prime\\\\
\end{aligned} \label{equ:dual-lp-best-response-p2}
\end{equation}
\end{multicols}
\vspace*{-4mm}
\noindent Where $q^\prime$ is a vector with $\left| \mathcal{A} \right| +1$ dual variables.
%Since the game restricted to $\mathcal{A}$ for Player {\la} is a zero-sum game, a minimax strategy for Player {\rational} is a best-response among . %, and so we can assume that Player {\rational} wishes to best-respond to Player {\la}.
Given some strategy $y^\prime$ for Player {\la}, Player {\rational} maximizes utility among strategies that induce $\mathcal{A}$. This gives the following best-response LP for Player {\rational}:
%\vspace{-2mm}
\begin{equation}
\begin{aligned}
  \max_{x} & \quad x^T (Ay^\prime)  \\
& x^T E^T = e^T  \\
& x \geq 0  \\
% x_{a^\prime}^T h_{a^\prime}  - x_a^T h_a \leq -\epsilon & \quad \forall a \mathcal{A}, a^\prime \notin \mathcal{A}, a[k]\ne a^\prime[k]  \\
% x_{\hat{a}}^T h_{\hat{a}} - x_{a}^T h_{a} \leq \epsilon & \quad \forall a,a^\prime \in \mathcal{A}\\
& x^T H_{\lnot \mathcal{A}} - x^T H_{\mathcal{A}} \leq -\epsilon \\
& x^T G_{\mathcal{A}^{*}} = x^T G_{\mathcal{A}} %\leq \epsilon
\end{aligned} \label{equ:primal-lp-best-response-p1}
%\vspace{-1.5mm}
\end{equation}
% Where each pair of actions $a,a^*\in A_I,a\notin \mathcal{A},a^*\in \mathcal{A}$ for some $I$ correspond to some column index $j$ into $H_{\lnot \mathcal{A}}, H_{\mathcal{A}}$.  with
Where the last two constraints encode equations~\eqref{equ:dominated-actions} and~\eqref{equ:incentivized-actions}, respectively. Equation~\eqref{equ:dominated-actions} is encoded via $H$ matrices that have a column for each pair $a\in \mathcal{A},a'\notin \mathcal{A}$ which has the expected value under $x$ for $a$ in $\mathcal{A}$ and for $a'$ in $\lnot\mathcal{A}$. Equation~\eqref{equ:incentivized-actions} is encoded analogously with $G$ for each pair $a,a^*\in \mathcal{A}$.
The dual problem uses the unconstrained vectors $p,v$ and constrained vector $u$ %, with $p$ holding $1+\left|\mathcal{I}_1\right|$ variables, $u$ holding
and looks as follows
%\vspace{-2mm}
\begin{equation}
\begin{gathered}
 \min_{p,u,v} \quad e^T p - \epsilon \cdot u\\% + \epsilon \cdot v \\
 E^T p + (H_{\lnot \mathcal{A}} - H_{\mathcal{A}}) u + (G_{\mathcal{A}^{*}} - G_{\mathcal{A}}) v \geq A^\prime y^\prime \\
 u \geq 0
\end{gathered} \label{equ:dual-lp-best-response-p1}
%\vspace{-1.5mm}
\end{equation}
% As with the standard sequence form LP approach,
We can now merge the dual~(\ref{equ:dual-lp-best-response-p2}) with the constraints from the primal~(\ref{equ:primal-lp-best-response-p1}) to compute a solution where Player {\rational} chooses $x$, which she will choose to minimize the objective of~(\ref{equ:dual-lp-best-response-p2}), a minmax strategy:
%\vspace{-2mm}
\begin{equation}
\begin{aligned}
  \min_{x,q^\prime} \quad q^{\prime T} f^\prime \\
 q^{\prime T} F^\prime - x^T B^\prime & \geq 0 \\
- x^T E^T  & = -e^T  \\
 x  & \geq 0  \\
 x^T H_{\mathcal{A}} - x^T H_{\lnot \mathcal{A}} & \geq \epsilon \\
 x^T G_{\mathcal{A}} - x^T G_{\mathcal{A}^{*}} & = 0 %\geq -\epsilon
\end{aligned} \label{equ:primal-lp-equilibrium}
%\vspace{-1.5mm}
\end{equation}
Taking the dual of this gives
%\vspace{-2mm}
\begin{equation}
\begin{aligned}
  \max_{y^\prime, p} \quad  -e^Tp + \epsilon \cdot u \\ % - \epsilon \cdot v \\
  -E^T p + (H_{\mathcal{A}} - H_{\lnot \mathcal{A}}) u + (G_{\mathcal{A}} - G_{\mathcal{A}^{*}})v & \leq B^\prime y^\prime \\
  F^\prime y^\prime & = f^\prime \\
  y,u & \geq 0
\end{aligned} \label{equ:dual-lp-equilibrium}
%\vspace{-1mm}
\end{equation}
% These primal and dual LPs can be used to compute Nash equilibria in the strategy spaces defined by some choice of actions $\mathcal{A}$ to be made optimal according to $h$.
% Player~$1$ computing a solution that enforces some $\mathcal{A}$ and Player~$2$ playing in the space of $\mathcal{A}$.
We are now ready to prove Theorem~\ref{the:nash-equilibrium-induced-game}.
\begin{proof}
The LPs are~(\ref{equ:primal-lp-equilibrium}) and~(\ref{equ:dual-lp-equilibrium}). We will use duality to show that they provide optimal solutions to each of the best response LPs. Since $A=-B$, the first constraint in (\ref{equ:dual-lp-equilibrium}) can be multiplied by $-1$ to obtain the first constraint in~(\ref{equ:dual-lp-best-response-p1}) and the objective function can be transformed to that of~(\ref{equ:dual-lp-best-response-p1}) by making it a minimization. By the weak duality theorem, we get the following inequalities
\begin{align*}
  q^{\prime T} f^\prime \geq x^TB^\prime y^\prime  \text{; by LPs \eqref{equ:primal-lp-best-response-p2} and \eqref{equ:dual-lp-best-response-p2}} \\
  % e^Tp - \epsilon\cdot u + \epsilon \cdot v \geq x^T A^\prime y^\prime & \text{; by LPs \ref{equ:primal-lp-best-response-p1} and \ref{equ:dual-lp-best-response-p1}}
  e^Tp - \epsilon\cdot u \geq x^T A^\prime y^\prime  \text{; by LPs \eqref{equ:primal-lp-best-response-p1} and \eqref{equ:dual-lp-best-response-p1}}
%\Rightarrow     q^{\prime T} f^\prime \geq x^TB^\prime y^\prime =  -x^T A^\prime y^\prime \geq   -e^Tp + \epsilon\cdot u % - \epsilon \cdot v
%   \label{equ:primal-dual-inequalities}
\end{align*}
We can multiply the last inequality by $-1$ to get:
% Taken together, this gives
 \begin{equation}
   q^{\prime T} f^\prime \geq x^TB^\prime y^\prime =  -x^T A^\prime y^\prime \geq   -e^Tp + \epsilon\cdot u % - \epsilon \cdot v
   \label{equ:primal-dual-inequalities}
 \end{equation}
By the strong duality theorem, for optimal solutions to LPs~\eqref{equ:primal-lp-equilibrium} and~\eqref{equ:dual-lp-equilibrium} we have equality in the objective functions
$
  q^{\prime T} f^\prime = -e^Tp + \epsilon u
$
which yields equality for Equation~\eqref{equ:primal-dual-inequalities}, and thereby equality for the objective functions in LPs~\eqref{equ:primal-lp-best-response-p2}, \eqref{equ:dual-lp-best-response-p2} and for \eqref{equ:primal-lp-best-response-p1}, \eqref{equ:dual-lp-best-response-p1}. By strong duality, this implies that any primal solution $x,q^\prime$ and dual solution $y^\prime,p$ to LPs~\eqref{equ:primal-lp-equilibrium} and~\eqref{equ:dual-lp-equilibrium} yields optimal solutions to the LPs~\eqref{equ:primal-lp-best-response-p2} and~\eqref{equ:primal-lp-best-response-p1}. Both players are thus best responding to the strategy of the other agent, yielding a Nash equilibrium.

Conversely, any Nash equilibrium gives optimal solutions $x,y^\prime$ for LPs~\eqref{equ:primal-lp-best-response-p2} and~\eqref{equ:primal-lp-best-response-p1}. With corresponding dual solutions $p,q^\prime$, equality is achieved in Equation~\eqref{equ:primal-dual-inequalities}, meaning that LPs~\eqref{equ:primal-lp-equilibrium} and~\eqref{equ:dual-lp-equilibrium} are solved optimally.

It remains to show the size bound for LP~\eqref{equ:primal-lp-equilibrium}. Using sparse representation, the number of non-zero entries in the matrices $A,B,E,F$ is linear in the size of the game tree.

The constraint set
$%\begin{equation*}
   x^T H_{\mathcal{A}} - x^T H_{\lnot \mathcal{A}}  \geq \epsilon
$,  when naively implemented, is not. The value of a deactivated sequence at some information set $I$ is dependent on the choice among the cartesian  product of choices at each information set $I^\prime$ encountered in hypothetical play below it. In practice we can avoid this by having a real-valued variable $v^d_{I}(I^\prime)$ representing the value of $I^\prime$ in lookahead from $I$, and introducing constraints
 \[
 v^d_{I}(I^\prime) \geq v^d_{I}(I^\prime,a)
 \]
 for each $a\in I^\prime$, where $v^d_{I}(I^\prime,a)$ is a variable representing the value of taking $a$ at $I^\prime$. If there are more information sets below $I^\prime$ where Player {\la} plays, before the lookahead depth is reached, we recursively constrain  $v^d_{I}(I^\prime,a)$ to be:
 \begin{equation}
v^d_{I}(I^\prime,a)\geq  \sum_{\check{I}\in \mathcal{D}}v^d_I(\check{I})
\end{equation}
where $\mathcal{D}$ is the set of information sets at the next level where Player {\la} plays. If there are no more information sets where Player {\la} acts, then we constrain $v^d_{I}(I^\prime,a)$:
 \begin{equation}
v^d_{I}(I^\prime,a)\geq  \sum_{s \in S_{I^\prime,a}^k} \pi^\sigma_{-\la} h(s)
\end{equation}
% where this expression is either a sum over $v_{\hat{I}}$ for information sets at the next level where the limited-lookahead player plays, or a sum over heuristic values of nodes if the lookahead depth is reached before the player plays again.
Setting it to the probability-weighted heuristic value of the nodes reached below it.

Using this, we can now write the constraint that $a$ dominates all $a^\prime\in I, a^\prime\notin \mathcal{A}$ as:
\[
\sum_{s\in S_{I,a}^k} \pi^\sigma(s) h(s) \geq v^d_I(I)
\]

 There can at most be $O(\sum_{I\in \mathcal{I}_{\la}} \left|A_I\right|)$ actions to be made dominant. For each action at some information set $I$, there can be at most $O(\max_{s \in S} \left| A_s \right|^{\min\{k,k^\prime\}})$ entries over all the constraints, where $k^\prime$ is the maximum depth of the subtrees rooted at $I$. This is because each node at the depth the player looks ahead to has its heuristic value added to at most one expression.
 % , since this value upper bounds the size of the subtree given by the root $\child{s}{a}$ and looking ahead to depth $k$.

 For the constraint set
$
 x^T G_{\mathcal{A}} - x^T G_{\mathcal{A}^{*}} = 0 %\geq -\epsilon
 $,
 the choice of hypothetical plays has already been made for both expressions, and so we have the constraint
 \[
\sum_{s \in S_{I,a}^k} \pi^{\sigma}_{}(s) h(s) = \sum_{s \in S_{I,a^\prime}^k} \pi^{\sigma^\prime}_{}(s) h(s)
\]
for all $ I\in\mathcal{I}_{\la}, a,a^\prime \in I,  \{a,\sigma^{\la}\},\{a^\prime,\sigma^{\la,\prime}\} \in \mathcal{A}$, where
\[
\sigma=\{\sigma_{-{\la}},\sigma^{\la}\}, \sigma^\prime=\{\sigma_{-{\la}},\sigma^{\la,\prime}\}
\]
There can at most be $\sum_{I\in \mathcal{I}_{\la}} \left|A_I\right|^2$ such constraints. Which is dominated by the size of the previous constraint set.

Summing up gives the desired bound.%, we get that the size of the LP is
% \[
% O(\left|S\right| +
% \sum_{I\in \mathcal{I}_{\la}} \left|A_I\right| \cdot \max_{s \in S} \left| A_s \right|^{\min\{k,k^\prime\}})
% \]
% O(\max_{I\in\mathcal{I}_\la} \left|A_I\right|^2 \cdot \left|\mathcal{I}_\la \right| \cdot \min\{\left|S\right|, \max_{s \in S} \left| A_s \right|^k\})
% \]
\end{proof}

In reality we are not given $\mathcal{A}$. To find a commitment strategy for Player {\rational}, we could loop through all possible structures $\mathcal{A}$, solve LP~\eqref{equ:primal-lp-equilibrium} for each one, and select the one that gives the highest value.

We show that this can be done without such exhaustive enumeration.  We introduce a MIP formulation that picks the optimal induced game $\mathcal{A}$. The MIP is given in~(\ref{equ:mip-adversarial}). We introduce Boolean sequence-form variables  %for some $a\in I \in \mathcal{I}_{\la}$
that denote making sequences suboptimal choices. These variables are then used to deactivate subsets of constraints, so that the MIP branches on formulations of LP~\eqref{equ:primal-lp-equilibrium}, that is, what goes into the structure $\mathcal{A}$. The size of the MIP is of the same order as that of LP~\eqref{equ:primal-lp-equilibrium}.
\begin{equation}
\begin{aligned}
  \min_{x,q,z} \quad q^{T} f \\
    q^TF &\geq x^TB - z M\\
    Ex  & = e  \\
  x^T H_{\mathcal{A}}  &\geq x^T H_{\lnot \mathcal{A}} + \epsilon - (1-z) M \\
 x^T G_{\mathcal{A}} & = x^T G_{\mathcal{A}^{*}}  \pm (1-z) M \\
\sum_{a\in A_I} z_a & \geq z_{a^\prime} \\
x  \geq 0, &\quad z \in \{0,1\}
\end{aligned} \label{equ:mip-adversarial}
%\vspace{-1.5mm}
\end{equation}

%\end{equation}
The variable vector $x$ contains the sequence form variables for Player {\rational}. The vector $q$ is the set of dual variables for Player {\la}.
$z$ is a vector of Boolean variables, one for each Player \la\ sequence. Setting $z_a=1$ denotes making the sequence $a$  an inoptimal choice. The matrix $M$ is a diagonal matrix with sufficiently large constants (for example, the smallest value in $B$) such that setting $z_a=1$ deactivates the corresponding constraint. %For all entries in $y_b$ set to active, the corresponding constraint becomes deactivated. %For a given sequence, the corresponding entry in $M$ can be set to the smallest value in $B$ belonging to a leaf node under the sequence.
Similar to the favorable-lookahead case, we introduce sequence form constraints $\sum_{a\in A_I} z_a \geq z_{a^\prime}$ where $a^\prime$ is the parent sequence,  to ensure that at least one action is picked when the parent sequence is active.
%For any sequence $a\in \mathcal{A}$, we must ensure that the constraints are inactive for all descendant sequences, as these are unreachable. This is done by introducing % the following constraint for all such $a$ and descendant sequences $a^\prime$
%the constraint
%$z_a \leq z_{a^\prime}$
%for all such $a$ and descendant sequences $a^\prime$.
% Similarly, the constraints
% \begin{equation*}
%   x^T H_{\mathcal{A}} - x^T H_{\lnot \mathcal{A}}  \geq \epsilon
% \end{equation*}
% \begin{equation*}
%  x^T G_{\mathcal{A}} - x^T G \geq -\epsilon
% \end{equation*}
% Are modified to
We must also ensure that the incentivization constraints are only active for actions in $\mathcal{A}$:
\begin{align}
  x^T H_{\mathcal{A}} - x^T H_{\lnot \mathcal{A}}  \geq \epsilon - (1-z) M \label{equ:appendix-incentive-constraint}\\
 x^T G_{\mathcal{A}} - x^T G_{\mathcal{A}^{*}} = 0 \pm (1-z) M \nonumber
\end{align}
For diagonal matrices $M$ with sufficiently large entries. Equality is implemented with a pair of inequality constraints. The $\pm$ denotes adding or subtracting, respectively, for the two inequalities.

The values of each column constraint in equation~\eqref{equ:appendix-incentive-constraint} is implemented by a series of constraints. We add Boolean variables $\sigma^I_{\la}(I^\prime,a^\prime)$ for each information set action pair $I^\prime,a^\prime$ that is potentially chosen in hypothetical play at $I$. Using our regular notation, for each $a,a^\prime$ where $a$ is the action to be made dominant, the constraint is implemented by:
\begin{equation}
  \sum_{s \in S_{I,a}^k} v^i(s) \geq v^d_I(I),\quad v^i(s) \leq \sigma^I_{\la}(I^\prime,a^\prime) \cdot M
\end{equation}
where the latter ensures that $v^i(s)$ is only non-zero if chosen in hypothetical play. We further need the constraint $v^i(s) \leq \pi_{-\la}^\sigma(s)h(s) $ to ensure that $v^i(s)$, for a node $s$ at the lookahead depth, is at most the heuristic value weighted by the probability of reaching $s$.

Since we have just modified existing constraints, and added variables and entries corresponding to the number of sequences and information sets, the size of this MIP has size on the order of the size of LP~\eqref{equ:primal-lp-equilibrium}.
% \[
% O(\left|S\right|) +
% O(\sum_{I\in \mathcal{I}_{\la}} \left|A_I\right| \cdot \max_{s \in S} \left| A_s \right|^{\min\{k,k^\prime\}})
% \]

\section{Experiments}
In this section we experimentally investigate how much utility can be gained by optimally exploiting a limited-lookahead player. We take a conservative approach, and assume that ties are broken adversarially. We conduct experiments on Kuhn poker~\cite{Kuhn50:Simplified}, a canonical testbed for game-theoretic algorithms, %and one of the first games to be studied game theoretically~\cite{cite nash here, but make sure it is the right citation, or should it be kuhn or vonNeumann-Morgenstern??? perhaps cite all three},
and a larger simplified poker game that we call {\KJ}. % The deck in {\KJ} consists of two kings and two jacks, and has two betting rounds. A full description of the two games is given in the appendix.

Kuhn poker consists of a three-card deck: king, queen, and jack. %Each player antes $1$, and is dealt a private card from the deck. (1) Player $1$ can then check or raise by $1$. If Player $1$ checks, (2) Player $2$ can check or raise by $1$. If Player $2$ checks, there is a showdown. If Player $2$ raises, (3) Player $1$ can call or fold. If Player $1$ calls, there is a showdown. If Player $1$ folds, Player $2$ gets the pot. If Player $1$ raises at (1), Player $2$ can either call or fold. If Player $2$ calls, there is a showdown. If Player $2$ folds, Player $1$ wins the pot.
Each player antes 1.
Each player is then dealt one of the three cards, and the third is put aside unseen. A round of betting occurs:
\begin{compactitem}
\item Player $1$ can check or bet 1.
  \begin{compactitem}
  \item If Player $1$ checks Player $2$ can check or raise 1.
    \begin{compactitem}
      \item If Player $2$ checks there is a showdown.
      \item If Player $2$ raises Player $1$ can fold or call.
        \begin{compactitem}
          \item If Player $1$ folds Player $2$ takes the pot.
          \item If Player $1$ calls there is a showdown for the pot.
          \end{compactitem}
        \end{compactitem}
      \item If Player $1$ raises Player $2$ can fold or call.
        \begin{compactitem}
        \item If Player $2$ folds Player $1$ takes the pot.
        \item If Player $2$ calls there is a showdown.
        \end{compactitem}
      \end{compactitem}
    \end{compactitem}

\noindent In a showdown, the player with the higher card wins the pot.

% In {\KJ}, the deck consists of two kings and two jacks. Each player initially
% puts $1$ chip in the pot, and is dealt a private card from the deck. A round of betting occurs. If neither player folded during betting, a public card is drawn from the remaining deck, and another round of betting occurs. If neither player folds during this round of betting, a showdown for the pot occurs.
In {\KJ}, the deck consists of two kings and two jacks. Each player antes $1$.
A private card is dealt to each, followed by a betting round ($p=2$), then a public card is dealt, followed by another betting round ($p=4$). If no player has folded, a showdown occurs.
Each round of betting looks as follows:
\begin{compactitem}
\item Player $1$ can check or bet $p$.
  \begin{compactitem}
  \item If Player $1$ checks Player $2$ can check or raise $p$.
    \begin{compactitem}
      \item If Player $2$ checks the betting round ends.
      \item If Player $2$ raises Player $1$ can fold or call.
        \begin{compactitem}
          \item If Player $1$ folds Player $2$ takes the pot.
          \item If Player $1$ calls the betting round ends.
          \end{compactitem}
        \end{compactitem}
      \item If Player $1$ raises Player $2$ can fold or call.
        \begin{compactitem}
        \item If Player $2$ folds Player $1$ takes the pot.
        \item If Player $2$ calls the betting round ends.
        \end{compactitem}
      \end{compactitem}
\end{compactitem}
% For each round of betting, Player $1$ has the option of raising or checking. If Player $1$ raised, Player $2$ can either call or fold. If Player $1$ checked, Player $2$ can
% either raise or check, and if Player $2$ raised, Player $1$ can
% finally call or fold. After the final round of betting, if neither
% player has folded, a showdown occurs.
Showdowns have two possible
outcomes: One player has a pair, or both players have the same private
card. For the former, the player with the pair wins the pot. For the
latter the pot is split.

% In {\KJ}, the deck consists of two kings and two jacks. Each player initially
% puts $1$ chip in the pot, and is dealt a private card from the deck. A round of betting occurs. If neither player folded during betting, a public card is drawn from the remaining deck, and another round of betting occurs. For each round of betting, Player $1$ has the option of raising or checking. If Player $1$ raised, Player $2$ can either call or fold. If Player $1$ checked, Player $2$ can
% either raise or check, and if Player $2$ raised, Player $1$ can
% finally call or fold. After the final round of betting, if neither
% player has folded, a showdown occurs. Showdowns have two possible
% outcomes: One player has a pair, or both players have the same private
% card. For the former, the player with the pair wins the pot. For the
% latter the pot is split.
Kuhn poker has $55$ nodes in the game tree and $13$ sequences per player. The {\KJ} game tree has $199$ nodes, and $57$ sequences per player.
All Kuhn instances solve in less than 0.2 seconds. Most KJ instances solve in less than 2 seconds, with a few lookahead~2 instances taking about 10 seconds. We also tried our MIP on the larger ``Leduc'' poker game (which has 1935 nodes in the game tree), but there the MIP did not solve within 2 hours.

To investigate the value that can be derived from exploiting a limited-lookahead opponent, a node evaluation heuristic is needed.
% There are obviously many different potential evaluation functions.
In this work we consider heuristics derived from a Nash equilibrium. For a given node, the heuristic value of the node is simply the expected value of the node in (some chosen) equilibrium. %, potentially with noise added.
This is arguably a conservative class of heuristics, as a limited-lookahead opponent would not be expected to know the value of the nodes in equilibrium.
% In fact, one might say that such knowledge violates the idea of limited-lookahead.
Even with this form of evaluation heuristic it is possible to exploit the limited-lookahead player, as we will show. We will also consider Gaussian noise being added to the node evaluation heuristic, more realistically modeling opponents who have vague ideas of the values of nodes in the game.
Formally, let $\sigma$ be an equilibrium, and $i$ the limited-lookahead player. The heuristic value $h(s)$ of a node $s$ is:
\begin{align}
  %\vspace{-2mm}
  h(s) = \begin{cases}
    u_i(s)  & \mbox{if } s\in Z\\
   \sum_{a\in A_s} \sigma(s,a) h(\child{s}{a})  & \mbox{otherwise} \\
 \end{cases}
 %\vspace{-1mm}
\end{align}
We consider two different noise models. The first adds Gaussian noise with mean $0$ and standard deviation $\gamma$  independently to each node evaluation, including leaf nodes. Letting $\mu_s$ be a noise term drawn i.i.d from $\mathcal{N}(0,\gamma)$: $\hat{h}(s) = h(s) + \mu_s$.
% \begin{align}
%   %\vspace{-2mm}
%   \hat{h}(s) = h(s) + \mu_s%\mathcal{N}(0,\gamma^2)
%   %\vspace{-2mm}
% \end{align}
The second, more realistic, model adds error cumulatively, with no error on leaf nodes:
\begin{align}
  %\vspace{-2mm}
  \bar{h}(s) = \begin{cases}
    u_i(s)  & \mbox{if } s\in Z\\
   \left[ \sum_{a\in A_s} \sigma(s,a) \bar{h}(\child{s}{a}) \right] +\mu_s  & \mbox{otherwise} \\
 \end{cases}
 %\vspace{-2mm}
\end{align}

\begin{figure}[!h]
  %\vspace{-15mm}
  \centering
  \includegraphics[scale=0.5]{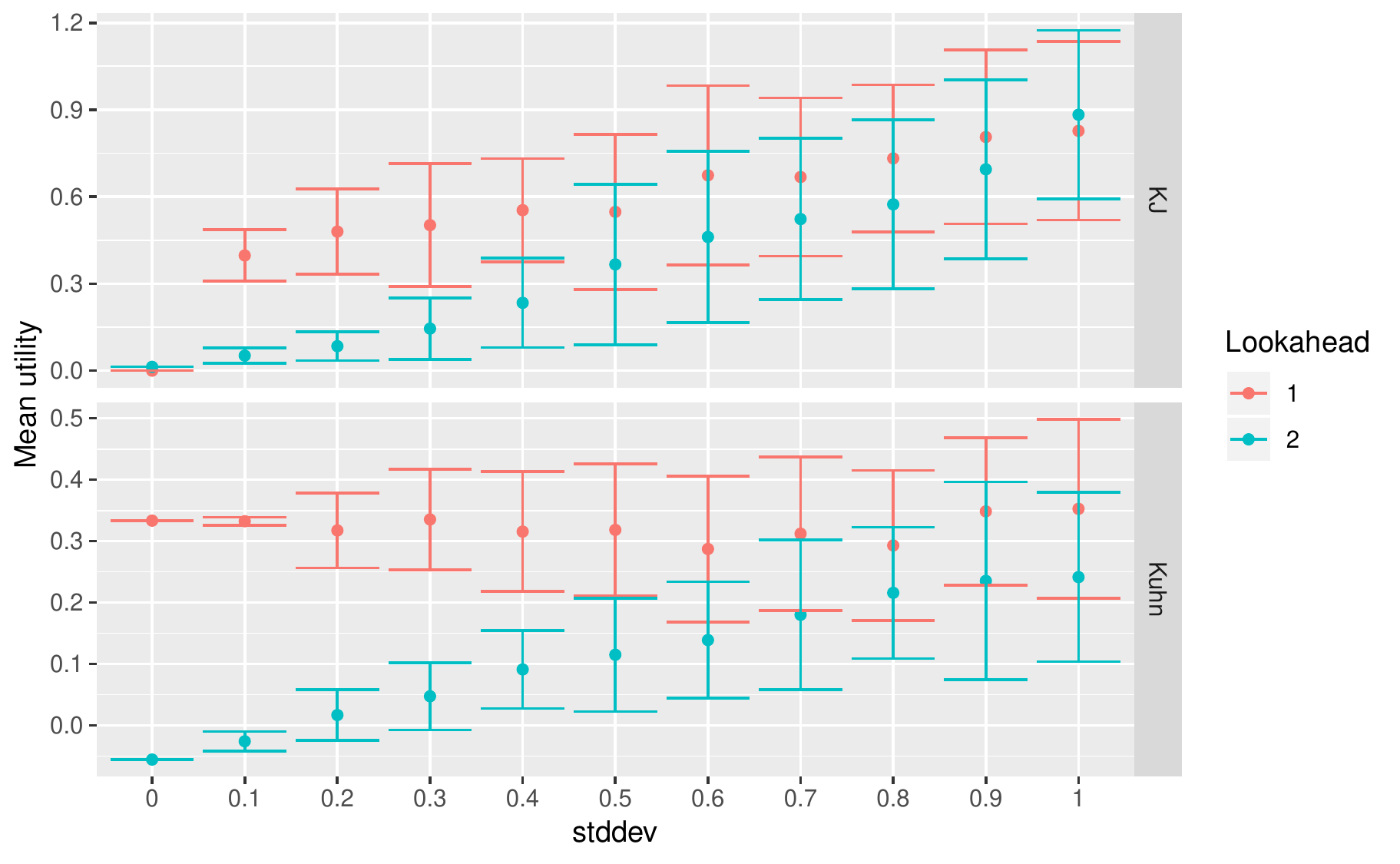}
    % \subfloat[Kuhn Player $1$, $\hat{h}$]{\includegraphics[scale=0.31]{figures/kuhnp1.pdf}}
    % \subfloat[Kuhn Player $2$, $\hat{h}$]{\includegraphics[scale=0.31]{figures/kuhnp2.pdf}}\\
    % \vspace{-0.55in}
    % \subfloat[KJ Player $1$, $\hat{h}$]{\includegraphics[scale=0.31]{figures/leduckj1raisep1.pdf}}
    % \subfloat[KJ Player $2$, $\hat{h}$]{\includegraphics[scale=0.31]{figures/leduckj1raisep2.pdf}}\\
    %\vspace{-3mm}
  \caption{Winnings in Kuhn poker and KJ for  Player $1$ being a rational player and Player $2$ having limited lookahead, respectively, for varying evaluation function noise in the evaluation function $\hat{h}$.}
  \label{fig:exploiting_hat}
%\vspace{-4mm}
\end{figure}

\begin{figure}[!h]
  %\vspace{-15mm}
  \centering
  \includegraphics[scale=0.5]{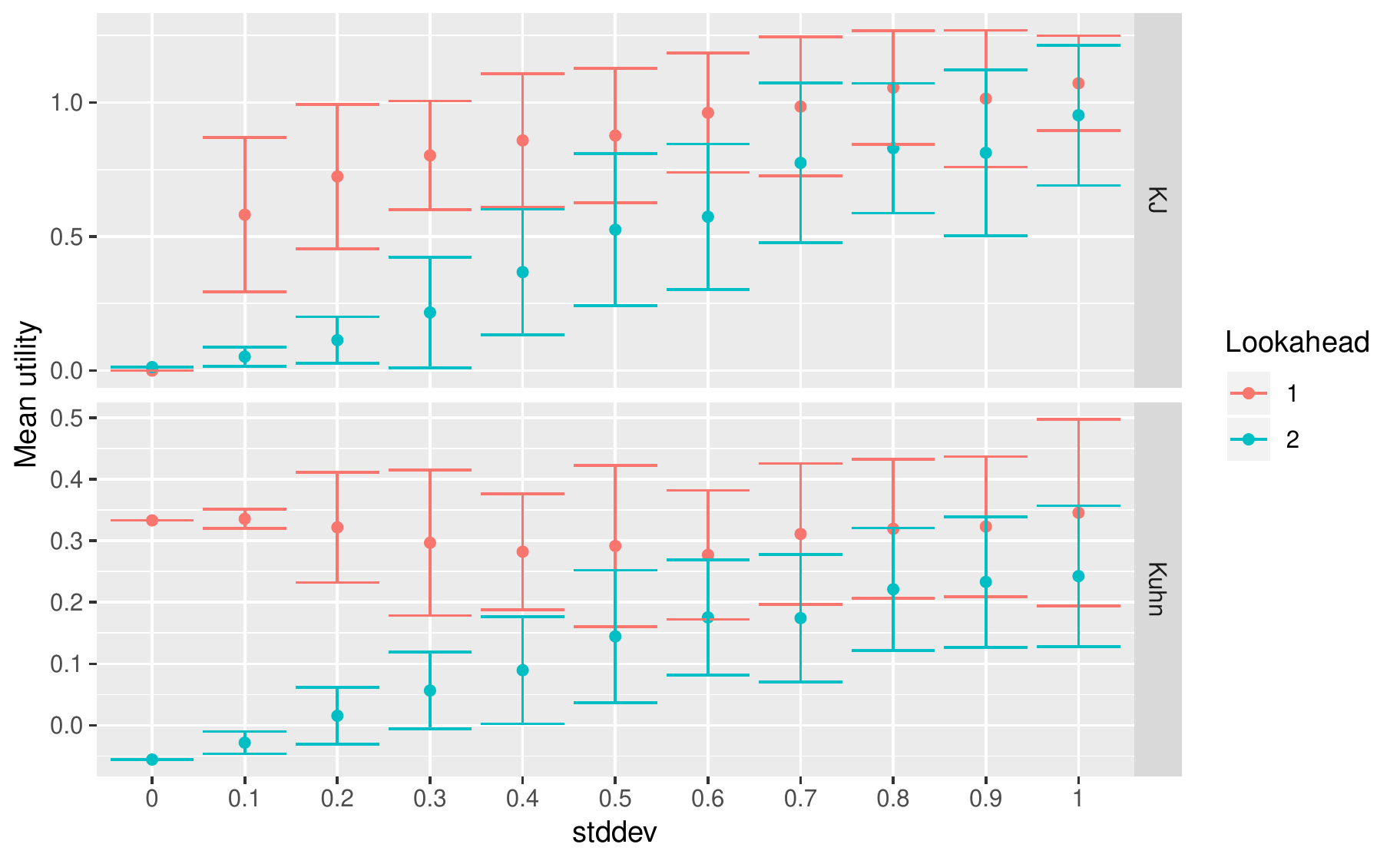}
    % \subfloat[Kuhn Player $1$, $\bar{h}$]{\includegraphics[scale=0.31]{figures/cumulativekuhnp1.pdf}}
    % \subfloat[Kuhn Player $2$, $\bar{h}$]{\includegraphics[scale=0.31]{figures/cumulativekuhnp2.pdf}}\\
    % \vspace{-0.55in}
    % \subfloat[KJ Player $1$, $\bar{h}$]{\includegraphics[scale=0.31]{figures/cumulativeleduckj1raisep1.pdf}}
    % \subfloat[KJ Player $2$, $\bar{h}$]{\includegraphics[scale=0.31]{figures/cumulativeleduckj1raisep2.pdf}}
    %\vspace{-3mm}
  \caption{Winnings in Kuhn poker and KJ for  Player $1$ being a rational player and Player $2$ having limited lookahead, respectively, for varying evaluation function noise in the evaluation function $\bar{h}$.}
  \label{fig:exploiting_bar}
%\vspace{-4mm}
\end{figure}

Using the MIP described in the Algorithms section, we computed optimal strategies for the rational player in Kuhn poker and {\KJ}. The MIP models were solved by CPLEX version 12.5. The results are given in Figures~\ref{fig:exploiting_hat} and~\ref{fig:exploiting_bar}. The x-axis is the noise parameter $\gamma$ for the standard deviation in $\hat{h}$ and $\bar{h}$. The y-axis is the corresponding utility for the rational player, averaged over at least 1000 runs for each tuple $\langle$game, choice of rational player, lookahead, standard deviation$\rangle$.
Each figure contains plots for the limited-lookahead player having lookahead $1$ or $2$.
% , and a baseline for the value of the game in equilibrium without a limit on lookahead
%The value of {\KJ} in equilibrium is $0$, so the baseline coincides with the x-axis for some of the plots.
At each point, the error bars show the standard deviation.

Figure~\ref{fig:exploiting_hat} (bottom) show the results for using evaluation function $\hat{h}$ in Kuhn poker, with the rational player being Player $1$. For rational Player $1$, we see that, even with no noise in the heuristic (that is, the limited-lookahead player knows the value of each node in equilibrium), it is possible to exploit the limited-lookahead player if she has lookahead depth 1. (With lookahead 2 she achieves the value of the game.) For either player and both amounts of lookahead, the exploitation potential steadily increases as noise is added.

  Figure~\ref{fig:exploiting_hat} (top) show the same variant for {\KJ}. Here, lookahead~1 is significantly worse than lookahead~2 for low amounts of noise. However, as more noise is added they become about the same.

  % lookahead~2 is actually worse for the limited-lookahead player than lookahead~1. To our knowledge, this is the first known imperfect-information {\em lookahead pathology}. Such pathologies have long been known in perfect-information games~\cite{Beal80:Analysis,Pearl81:Heuristic,Nau83:Pathology}, and understanding them remains an active area of research~\cite{Luvstrek06:Real,Nau10:When,Wilson12:Improving}. This version of the node heuristic does not have increasing {\em visibility}: node evaluations do not get more accurate toward the end of the game. Our experiments on {\KJ} with $\bar{h}$ in Figures~\ref{fig:exploiting_bar} c and d do not have this pathology, and $\bar{h}$ does have increasing visibility.

% Now, assuming that P2 is the limited-lookahead player and minimizing, for large enough $\alpha$, the node labeled P$1^*$ will be more desirable than any other node in the game, since it has expected value $-\alpha$ according to the evaluation function. A rational player P$1$ can use this to get P$2$ to go down at P$2^*$, and then switch to the action that leads to $\alpha$. This example is for lookahead $1$, but we can generalize the example to work with any finite lookahead depth: the node P$1^*$ can be replaced by a subtree where every other leaf has payoff $2\alpha$, in which case P$2$ would be forced to go to the leaf with payoff $\alpha$ once down has been chosen at P$2^*$.

Figure~\ref{fig:exploiting_bar} (bottom) shows the results for Kuhn with $\bar{h}$. The performance is very similar to the results for $\hat{h}$, with almost identical expected utility for all scenarios. Figure~\ref{fig:exploiting_bar} (top), as previously mentioned, shows the results with $\bar{h}$ on {\KJ}. Here we see no abstraction pathologies, and for the setting where Player $2$ is the rational player we see the most pronounced difference in exploitability based on lookahead.

\section{Conclusions and future research}
In this paper, we initiated the study of limited lookahead in imperfect-infor\-mation games. As a generalization of limited lookahead in perfect-information games, we find it interesting in its own right.  The game-theoretic reasoning over limited lookahead is another novel aspect.  The model also has applications, for example in security games and in steering evolution/adaptation in biomedical games.

We characterized the complexity of finding a Nash equilibrium and optimal strategy to commit to for either player. Figure~\ref{fig:complexity-overview} summarized those results.
% \begin{figure}[h]
%   %\vspace{-2mm}
%   \centering
%         \scalebox{1.25}{
%         \input{figures/complexity-tree}
%     }
%   %\vspace{-3mm}
%   \caption{Our complexity results. \{PPAD,NP\}-hard indicates that finding a Nash equilibrium is PPAD-hard and finding an optimal strategy to commit to is NP-hard. P indicates polynomial time.
%   }
%   \label{fig:complexity-overview}
%   %\vspace{-2mm}
% \end{figure}

We then designed several MIPs for computing optimal strategies to commit to for the rational player in the general NP-hard cases. First, we showed that the sequence form can be used to design a MIP that has size almost linear in the size of the game tree for many practical games, when ties are broken statically or in favor of the rational player. We then showed that when ties are broken adversarially, the problem reduces to choosing the best among a set of two-player zero-sum games (the tie-breaking being the opponent), and for each of those games the optimal strategy can be computed with an LP. We then introduced a MIP formulation that branches on these games to find the optimal solution.

We experimentally studied the impact of limited lookahead in two poker games. We
demonstrated that it is possible to achieve large utility gains by exploiting a
limited-lookahead opponent. As one would expect, the limited-lookahead player
often obtains the value of the game if her heuristic node evaluation is exact
(that is, it gives the expected values of nodes in the game tree for some
equilibrium)---but we provided a counterexample that shows that this is not
sufficient in general. Finally, we studied the impact of noise in those
estimates, and different lookahead depths.
% While lookahead 2 outperformed
% lookahead 1 in our experiments, we uncovered an imperfect-information game
% lookahead pathology: deeper lookahead can hurt the limited-lookahead player. We
% demonstrated how this can occur with any finite depth of lookahead, even if the
% limited-lookahead player's node evaluation heuristic returns exact values from
% an equilibrium.

Our algorithms in the NP-hard adversarial tie-breaking setting scaled to games
with hundreds of nodes. For some practical settings, significantly more
scalability will be needed. There are at least two exciting future directions
toward achieving this. One is to design faster---optimal or
good-enough---algorithms. The other is designing abstraction techniques for the
limited-lookahead setting. The latter could be used with our current algorithms,
or in conjunction with faster future algorithms. In extensive-form game solving
with rational players, abstraction plays an important role in large-scale game
solving~\citep{Sandholm15:Abstraction} and
theoretical solution quality guarantees have recently been
achieved~\cite{Lanctot12:No,Kroer18:Unified}.
Limited-lookahead games have much stronger structure, especially locally around
an information set, and it may be possible to utilize that to develop
abstraction techniques with significantly stronger solution quality bounds.
Also, leading practical game abstraction
algorithms~\citep{Ganzfried14:Potential,Brown15:Hierarchical}, while lacking
theoretical guarantees, could immediately be used to investigate exploitation
potential in larger games. One option would be to only perform lossy abstraction
on the leader's action space (which is often far larger than that of the
attacker), optionally followed by lossless abstraction on the attacker. If done
carefully, we may be able to ensure that we are correctly reasoning about the
follower's best response. In that case, the error from abstracting would give us
bounds on how close we are to the optimal strategy to commit to.
%
% It would also be interesting to explore conditions under which lookahead pathologies occur, and map out similarities and dissimilarities to the pathologies in perfect-information games.
%

Finally, uncertainty over $h$ is an important future research direction. This would lead to more robust solution concepts, thereby alleviating the pitfalls involved with using an imperfect estimate of $h$. In a follow-up conference paper~\citep{Kroer18:Robust} we show that it is indeed possible to handle robust models of the present work. In particular we build on the limited-lookahead model, as well as general Stackelberg MIP ideas from the present paper to construct MIPs that handle the robust case. There are still many interesting open questions on handling evaluation uncertainty. For example, can we construct tractable stochastic (rather than worst-case robust) uncertainty models?

\section{Acknowledgments}

This material is based on work supported by the National Science Foundation under grants IIS-1718457, IIS-1617590, IIS-1901403, CCF-1733556, IIS-1320620, and the ARO under award W911NF-17-1-0082.

\section*{References}
\bibliographystyle{elsarticle-harv}
\bibliography{../../dairefs/dairefs.bib}

\end{document}

%% file: figures/complexity-tree.tex
\begin{tikzpicture}
%\tikzset{level distance=15pt}
%\tikzset{execute at begin node=\strut}
%\tikzset{every tree node/.style={anchor=base west}}
\tikzstyle{level 1}=[sibling distance=-60pt]
\tikzstyle{level 2}=[sibling distance=0pt]
\tikzstyle{level 3}=[sibling distance=5pt]
\Tree [.{Information sets}
    \edge node[auto=right,scale=0.7]{no};
    [.{Lookahead depth $>1$}
        \edge node[auto=right,scale=0.7]{yes};
        {\{PPAD,NP\}-hard} 
        \edge node[auto=left,scale=0.7]{no};
        [.{Solution concept}
            \edge node[auto=right,scale=0.7]{Equilibrium};
            P
            \edge node[auto=left,scale=0.7]{Commitment}; 
            [.{Tie-breaking rule}
                \edge node[auto=right,scale=0.7]{Adversarial, static};
                P
                \edge node[auto=left,scale=0.7]{Favorable};
                {NP-hard}
                ]
            ]
        ]
    \edge node[auto=left,scale=0.7]{yes};
    [. {\{PPAD,NP\}-hard} 
    ]
]
\end{tikzpicture}

%% file: figures/3sat-break-ties-favorably.tex
\begin{tikzpicture}[level distance=1cm]
  \tikzstyle{every node}=[font=\small]
  \tikzstyle{p1 node}=[circle,draw,inner sep=1,fill=black]
  \tikzstyle{p2 node}=[circle,draw,inner sep=1]
  \tikzstyle{level 1}=[sibling distance=51mm]
  \tikzstyle{level 2}=[sibling distance=15mm]
  \tikzstyle{level 3}=[sibling distance=5mm]
  \node(naturenode){N} {
      child{node[p2 node](c1){}
          child{node[p1 node](0-0-0){}
              child{node(0-0-0-0){$1$}
                  edge from parent node[left]{$t$}
              }
              child{node(0-0-0-1){$0$}
                  edge from parent node[right]{$f$}
              }
              edge from parent node[left]{$v_1$\ \ }
          }
          child{node[p1 node](c1v2){}
              child{node(0-0-1-0){$0$}
                  edge from parent node[left]{$t$}
              }
              child{node(0-0-1-1){$1$}
                  edge from parent node[right]{$f$}
              }
              edge from parent node[left]{${\mathsmaller\lnot} v_2$}
          }
          child{node[p1 node](c1v3){}
              child{node(0-0-2-0){$0$}
                  edge from parent node[left]{$t$}
              }
              child{node(0-0-2-1){$1$}
                  edge from parent node[right]{$f$}
              }
              edge from parent node[left]{${\mathsmaller\lnot} v_3$}
          }
          edge from parent node[left](c1edge){$c_1$\ \ \ \ \ }
      }
      child{node[p2 node] (cm){}
          child{node[p1 node](cmv2){}
              child{node(0-0-0-0){$1$}
                  edge from parent node[left]{$t$}
              }
              child{node(0-0-0-1){$0$}
                  edge from parent node[right]{$f$}
              }
              edge from parent node[left]{$v_2$\ \ }
          }
          child{node[p1 node](cmv3){}
              child{node(0-0-1-0){$0$}
                  edge from parent node[left]{$t$}
              }
              child{node(0-0-1-1){$1$}
                  edge from parent node[right]{$f$}
              }
              edge from parent node[left]{${\mathsmaller\lnot} v_{3}$}
          }
          child{node[p1 node](cmv4){}
              child{node(0-0-2-0){$0$}
                  edge from parent node[left]{$t$}
              }
              child{node(0-0-2-1){$1$}
                  edge from parent node[right]{$f$}
              }
              edge from parent node[left]{${\mathsmaller\lnot} v_4$}
          }
          edge from parent node[right](cmedge){\ \ \ $c_m$}
      }
  };
  \draw [dashed, out=20, in=160](c1v2)to(cmv2);
  \draw [dashed, out=20, in=160](c1v3)to(cmv3);
  \coordinate (level12mid) at ($(c1edge)!0.5!(cmedge)$);
  % \draw[at level12mid]  node {...};
  \node[left of=cmedge](ellipsenode){...};
  % \draw [dashed](0-2)to(0-3);
  % \draw [dashed, out=20, in=160](0-0-0)to(0-2-0);
  % \draw [dashed, out=-20, in=-160](0-1-0)to(0-3-0);
\end{tikzpicture}

%% file: figures/3sat-look-ahead-2.tex
\begin{tikzpicture}[level distance=1cm]
  \tikzstyle{every node}=[font=\small]
  \tikzstyle{p1 node}=[circle,draw,inner sep=1,fill=black]
  \tikzstyle{p2 node}=[circle,draw,inner sep=1]
  \tikzstyle{level 1}=[sibling distance=15mm]
  \tikzstyle{level 2}=[sibling distance=5mm]
  \node(c1node){$c_1$} {
          child{node(0-0-0){$(0,\frac{2}{3})$}
              edge from parent node[left,pos=0.25]{$unsat$\ \ \ \ \ \ \ \ }
          }
          child{node[p1 node](0-0-0){}
              child{node(0-0-0-0){$1$}
                  edge from parent node[left]{$t$}
              }
              child{node(0-0-0-1){$0$}
                  edge from parent node[right]{$f$}
              }
              edge from parent node[left]{$v_1$}
          }
          child{node[p1 node](c1v2){}
              child{node(0-0-1-0){$0$}
                  edge from parent node[left]{$t$}
              }
              child{node(0-0-1-1){$1$}
                  edge from parent node[right]{$f$}
              }
              edge from parent node[left]{${\mathsmaller\lnot} v_2$}
          }
          child{node[p1 node](c1v3){}
              child{node(0-0-2-0){$0$}
                  edge from parent node[left]{$t$}
              }
              child{node(0-0-2-1){$1$}
                  edge from parent node[right]{$f$}
              }
              edge from parent node[right,pos=0.25]{\ \ ${\mathsmaller\lnot} v_3$}
          }
  };
  % \draw[at level12mid]  node {...};
  % \draw [dashed](0-2)to(0-3);
  % \draw [dashed, out=20, in=160](0-0-0)to(0-2-0);
  % \draw [dashed, out=-20, in=-160](0-1-0)to(0-3-0);
\end{tikzpicture}

%% file: figures/3sat-information-sets.tex
\begin{tikzpicture}[level distance=1cm]
  \tikzset{
    buffer/.style={
        draw,
        shape border rotate=90,
        isosceles triangle,
        isosceles triangle apex angle=30,
        minimum height=5.4em
    }
}
  \tikzstyle{every node}=[font=\small]
  \tikzstyle{p1 node}=[circle,draw,inner sep=1,fill=black]
  \tikzstyle{p2 node}=[circle,draw,inner sep=1]
  \tikzstyle{level 1}=[sibling distance=25mm]
  \tikzstyle{level 2}=[sibling distance=15mm]
  \tikzstyle{level 3}=[sibling distance=5mm]
  \node(naturenode){N} {
      child{node[p1 node](v1c1){}
          child{node[p2 node](0-0-0){}
              child{node(0-0-0-0){$1$}
                  edge from parent node[left]{$t$}
              }
              child{node(0-0-0-1){$0$}
                  edge from parent node[right]{$f$}
              }
              edge from parent node[left]{$t$}
          }
          child{node[p2 node](c1v3){}
              child{node(0-0-2-0){$0$}
                  edge from parent node[left]{$t$}
              }
              child{node(0-0-2-1){$1$}
                  edge from parent node[right]{$f$}
              }
              edge from parent node[left]{$f$}
          }
          edge from parent node[left,pos=0.25](v1c1edge){$v_1,c_1$\ \ \ \ }
      }
      child{node[p1 node] (v1cm){}
          edge from parent node[left,pos=0.7](v1cmedge){$v_1,c_m$}
      }
      child{node (vncm){1}
          edge from parent node[right,pos=0.25](vncmedge){\ \ \ \ \ $v_n,c_m$}
      }
  };
  \coordinate (level12mid) at ($(v1cmedge)!0.5!(vncmedge)$);
  % \draw[at level12mid]  node {...};
  \node[right of=v1cmedge](ellipsenode){...};
  
  \coordinate (t1c) at ($ (v1cm) - (0,1.6) $);
  \node[buffer] at (t1c) (triangle1){};
%  \coordinate (t2c) at ($ (vncm) - (0,1.6) $);
%  \node[buffer] at (t2c) (triangle2){};
  \draw [dashed](v1c1) to (v1cm);
  % \draw [dashed, out=20, in=160](0-0-0)to(0-2-0);
  % \draw [dashed, out=-20, in=-160](0-1-0)to(0-3-0);
\end{tikzpicture}